\documentclass[12pt,letterpaper]{report}

\usepackage[activeacute,english]{babel}
\usepackage{amsmath}
\usepackage{amsbsy}
\usepackage{amsthm}
\usepackage{amssymb}
\usepackage{latexsym}
\usepackage[dvips]{graphicx}

\begin{document}

\title{Instituto Superior de Ciencias y Tecnolog\'{\i}a Nucleares \\
\vspace{2cm} Dissertation Master of Science Thesis \\ \vspace{2.5cm}
{\bf Gluon, Quark and Hadron Masses from a \\ Modified
Perturbative QCD \\ \vspace{1.7cm}}}

\author{{\bf Author: Marcos Rigol Madrazo}
\\ \vspace{0.1cm} \and  {\bf Advisor: Dr. Alejandro Cabo Montes de Oca}}

\date{\vspace{1cm}Havana 2000}

\maketitle

\begin{abstract}

The development of a Modified Perturbation Theory for QCD,
introduced in previous works, is continued. The gluon propagator
is modified as consequence of a soft gluon pairs condensate in the
vacuum. The modified Feynman rules for $\alpha=1$ are shown, and
some physical magnitudes calculated with them. The mean value of
$G^{2}$, gluon masses and the effective potential are calculated
up to the $g^2$ order, improving previous calculations. In
connection with the gluon self-energy it follows that the gluonic
mass shell becomes tachyonic in the considered approximation. The
constituent quarks masses, produced by the influence of the
condensate, are also calculated. Results of the order of 1/3 of
the nucleon mass, are obtained for the constituent masses of the
up and down quarks. In addition, the predicted flavour dependence
of the calculated quarks masses turns out to be the appropriate to
reproduce the spectrum of the ground states within the various
groups of hadronic resonances, through the simple addition of
their constituent quark masses. Finally, the generalization of the
theory for arbitrary values of the $\alpha$ gauge parameter is
explored.

\end{abstract}

\tableofcontents

\chapter{Introduction}

The development of Quantum Chromodynamics (QCD) has been one of
the greatest achievements of the High Energy Theoretical Physics
in the latest thirty years. As it is well known the smallest of
the coupling constant at high momenta (asymptotic freedom), made
possible the development of the theory through the usual
perturbative language. That situation strongly simplifies the
study of such phenomena and the calculated results are in very
good agreement with the experimental data. However, this so called
perturbative QCD (PQCD) is far from being able to furnish even a
rough description for the relevant physics at low energies. The
solution of this situation is currently one of the main challenges
of the Theoretical Physics.

A relevant phenomenon related with QCD is the color confinement.
For the time being, there are strong reasons to believe that the
relation between the basic quantities in QCD, the colored gluon
and quarks fields, and the real world characterized by a whole
variety of colorless interacting mesons and baryons, can be
understood by solving the confinement problem. The basic picture
is that a theory in which $SU(3)$ is an exact symmetry, the
fundamental fields (quarks and gluons) cannot be associated to
physical states and then the true physical states consist only of
colorless mesons and baryons.

The above cited limitations of the PQCD means in particular that
the usual Fock space vacuum for the non-interacting theory, is
unable to predicts even approximately the real ground state of the
QCD \cite{Savvidy1}-\cite{Reuter}. This is at contrast with the
case in QED where the standard perturbation theory, based in the
empty of the Fock space, gives a more than good concordance with
the experimental results. Then, the nature of the vacuum structure
is one of the main problems to be clarified and naturally its
solution is closely related with the color confinement effect. A
review of the various models considered to investigate the
confinement and the vacuum structure in QCD can be found in Refs.
\cite{ShuryakTex,Shuryak2}.

Another important problem for QCD is the one related with quark
masses. There are no free quarks, so it is necessary to take care
with the meaning of quark masses. They should be understood as
mass parameters appearing in the QCD Lagrangian
\cite{Gell}-\cite{Weinberg1} (called current quark masses) and
their origins have generated polemic for long time
\cite{Weinberg2}. The theoretical ratios between the current quark
masses and their absolute values can be found with the standard
methods of current algebra. Their experimental values are
determined indirectly and the most recently reports for them
\cite{Report} will be used in the present work. On the other hand
the masses considered in phenomenological hadron models are called
constituent quark masses, which values are higher than the ones of
current masses, and it is supposed that such values come from
non-perturbative effects, like gluonic condensate effects for
example. Other properties of quarks, like magnetic moments,
couplings, and electro-weak form factors are not influenced by
this non-perturbative effects \cite{Weinberg34}.

In the former works \cite{1995}-\cite{tesis} an attempt to
construct a modified perturbation expansion for QCD, able to
predict some low energy properties of this theory, have been
considered. The proposed modified expansion \cite{1995} conserves
the color $SU(3)$ and Lorentz symmetries, and was implemented in
order to solve the symmetry limitations of the earlier
chromomagnetic field models \cite{Savvidy1}-\cite{Reuter} which
inspired the search. The similar lack of manifest Lorentz
invariance had also a modification of the Feynman rules,
attempting to include gluon condensation, advanced later
\cite{Hoyer} in which a delta function term, for $k < \Lambda
_{qcd}$, was considered in addition to the perturbative piece. The
expansion proposed in a previous work \cite{1995}, considers a
change of the gluon propagator in a term associated to a
condensation of zero momentum gluons. This change of the usual
rules had the interesting property of producing a non vanishing
value for the gluon condensation parameter $\langle g^2 G^2
\rangle$ \cite{Zakharov} already in the first approximations. In
addition, in Ref. \cite{1995} a non vanishing value for the
effective mass of the gluons was obtained at the one loop level.
Finally, a perturbative evaluation of the effective potential as a
function of the condensate parameter indicated that the
condensation is spontaneously generated from a zero condensate
state.

At this point it is important to remark that earlier works
\cite{Munczek,Burden} introduced a pure delta function at zero
momentum as gluon propagator, searching for a model of meson
resonances. In these cases the constants introduced, multiplying
the delta function, were different to the one fixed in our
previous work \cite{PRD,tesis}, which lead to also different
results. In particular, the singularity structure of the quark
propagator have no pole on the real $p^2$ axis \cite{Burden} at
difference with our results.

\newpage

Justifying the applicability of the Feynman expansion introduced
in Ref. \cite{1995} was advanced later \cite{PRD,tesis}. This was
in need because after modifying the propagator, it was unknown
whether or not the initial state generating the expansion was a
physical state of the theory. In Ref. \cite{PRD,tesis},
considering the operational formulation of QCD developed by Kugo
and Ojima \cite{Kugo,OjimaTex}, was possible to find a zero
momentum gluon condensate state able to generate through the Wick
expansion the sort of propagators considered in Ref. \cite{1995}.
It was checked that this condensate satisfies the physical state
conditions imposed by the BRST formalism. The discussion allowed
also to a more precise characterization of the class of changes
admitted in the diagramatic expansion by the physical state
conditions on the initial state. Specifically it was found that
the $C$ parameter, describing the gluon condensation, must be real
and positive. It should be remarked that an analogous proposal for
a zero mode gluon condensate state was independently advanced in
Ref. \cite{pavel}, in an attempt to give foundation to the
propagator proposed by Munczek and Nemirovski.

In the present work the development of the Modified Perturbation
Theory \cite{1995}-\cite{tesis} is continued, and the objectives
proposed are: a) To state clearly the details of the considered
perturbative expansion for $\alpha=1$. b) To realize an improved
evaluation of quantities calculated in Ref. \cite{1995}, that is,
the mean value of $G^{2}$, the gluon mass, and the effective
potential, by also imposing the restriction that appeared on the
parameter $C$ describing the condensation \cite{PRD,tesis}. c) To
apply the results of the calculation of $\langle G^2\rangle$ to
evaluate the effects of the condensate on the quark masses, after
fixing the parameter $g^{2}C$ to reproduce the accepted value for
$\langle g^2 G^2 \rangle$. d) Finally, to explore the
generalization of the theory for arbitrary values of the gauge
parameter $\alpha$.

The evaluations presented here are done in a special
approximation. It can be defined as the one allowed to be
considered with precision by the fact that in the present state of
the analysis the renormalization of the fields and coupling
constants has not being introduced yet. Therefore, the
calculations of the $\langle G^2\rangle$, the quark and gluon
masses will be searched not only up to the order $g^2$ but also
with the additional restriction of disregarding all the terms not
involving the condensate contribution. To take into account those
terms will need the introduction of the renormalization procedure
up to two loops, because the one loop divergence are unchanged by
the use of the new propagator. This is out of the scope of the
present exploration. It is worth noting that the just defined
approximation scheme resembles a sort of quasi-classical limit in
which the condensate would play the role of a macroscopic quantum
field. The correctness of such a picture is suggested by the
initial motivation of the analysis, as directed to construct a
covariant version of the chromomagnetic models \cite{1995}.

\newpage

The exposition will be organized as follows: Chapter 2 is divided
in two sections. In the first one a review of the previous work
\cite{PRD,tesis} is done. In the second section the Feynman Rules
for the considered perturbative expansion ($\alpha=1$) are stated.
Chapter 3 is divided in four sections. In the first one the mean
value of $G^2$ is calculated and the free parameter $g^2C$ fixed
from the accepted value of $\langle g^2G^2\rangle$. In the second
and third sections the gluon and quark masses respectively are
determined. In the four section the effective potential is
evaluate as a function of $\langle G^2\rangle$. Chapter 4 is
devoted to explore the generalization of the theory for arbitrary
values of the gauge parameter $\alpha$. Finally, two appendixes
are introduced. In the first one it is considered the exposition
of the diagrams calculated for determining the mean value of
$G^2$. The second one was introduced for an analysis of the most
elaborated parts in the calculation of longitudinal and scalar
modes contribution to the gluon propagator modification,
considered in Chapter 4.

\chapter{Modified Feynman Rules}

The Modified Feynman rules, advanced in previous works, are
stated. The gluon propagator is modified as consequence of a soft
gluon pairs condensate in the QCD vacuum, this fact is represented
by a delta function in momentum representation. In the first
section the previous work is shortly reviewed and in the second
section the Feynman rules exposed.

\section{Review of ``About an Alternative Vacuum State for
Perturbative QCD''}

In the previous work \cite{PRD,tesis} the construction of a
modified vacuum state for perturbative QCD was considered. By
using the operational formulation for the Quantum Gauge Field
Theory developed by Kugo and Ojima, a particular state vector for
QCD in the non-interacting limit, that obeys the BRST physical
state condition, was proposed. The general motivation for
searching this wave function was to give a foundation to the
perturbative expansion proposed in Ref. \cite{1995}. The proposed
state had the form

\begin{equation}
\mid \widetilde{\Phi }\rangle =\frac 1{\sqrt{N}}\exp
\sum\limits_a\left(\frac 12A_{0,1}^{a+}A_{0,1}^{a+}+\frac
12A_{0,2}^{a+}A_{0,2}^{a+}+B_0^{a+}A_0^{L,a+}+i\overline{c}
_0^{a+}c_0^{a+}\right) \mid 0\rangle. \label{Vacuum}
\end{equation}

The state (\ref{Vacuum}) is a coherent superposition of zero
momentum gluon pairs, expressed in terms of the transverse and
longitudinal gluon $ A_{0,1}^{a+}$, $A_{0,2}^{a+}$, $A_0^{L,a+}$,
ghosts $c_0^{a+}$, $\overline{c} _0^{a+}$, and Nakanishi-Lautrup
$B_0^{a+}$ creation operators of the free theory
\cite{Kugo,OjimaTex}. This state is colorless because the
contracted color index $a$. Eq. (\ref{Vacuum}) should be
understood in a sense of limit. The operators are supposed to be
evaluated at the smallest non-vanishing null momentum $\vec{p}$
resulting from quantizing the QCD free fields with periodic
boundary conditions in a large box of volume $V\rightarrow\infty$.
The coefficients of the products of creation operators are
slightly different from their shown numerical values before the
infinity volume limit is taken. They reach to the values in
expression (\ref{Vacuum}) when $V\rightarrow\infty \
(p\rightarrow0)$ in the way determined previously
\cite{PRD,tesis}.

For the construction of the state (\ref{Vacuum}), the BRST
physical state conditions
\begin{eqnarray}
&&Q_{B}\mid \Phi \rangle =0, \nonumber \\ &&Q_{C}\mid \Phi \rangle
=0, \label{fis}
\end{eqnarray}
\noindent were required, in which $Q_B$ and $Q_C$ are the charges
associated to the BRST symmetry and ghost number conservation. The
operational quantization was considered in a scheme of the
Gupta-Bleuler type, in which all the gluons are treated on the
same footing \cite{Kugo,OjimaTex}. This corresponds with the
selection of the gauge parameter $\alpha =1$ in the standard
functional approach. In justifying proceeding in such a way, the
central assumption was that under the adiabatic connection of the
color interaction, the evolution won't bring the state out of the
physical subspace at any stage of the connection. Thus, the final
state will be also a physical one for the interacting theory and
the associated Feynman expansion should have a physical meaning.
It is clear that this approach left out the question of the
construction of a gauge parameter independent formulation of the
theory. However, as a minimal logical ground for the physical
relevance of the predictions was given, the generalization of the
analysis for and arbitrary $\alpha$ was postponed. In Chapter 4 of
the present work this generalization is explored.

The gluon propagator corresponding to the modified vacuum state
(\ref{Vacuum}) was calculated \cite{PRD,tesis} through the
expression for the generating functional of the free particle
Green functions $Z_0\left(J\right)$ \cite{Gasiorowicz},
\begin{eqnarray}
&&\langle \widetilde{\Phi }\mid \exp \left\{ i\int
d^4x\sum\limits_{a=1,..,8}J^{\mu,a}\left(x\right) A_\mu
^{a-}\left(x\right) \right\} \exp \left\{ i\int
d^4x\sum\limits_{a=1,..,8}J^{\mu,a}\left(x\right) A_\mu
^{a+}\left(x\right) \right\} \mid \widetilde{\Phi } \rangle
\nonumber \\ &&\times \exp \left\{ \frac
i2\sum\limits_{a,b=1,..8}\int d^4xd^4yJ^{\mu,a}\left(x\right)
D_{\mu \nu }^{ab}(x-y)J^{\nu,b}\left(y\right) \right\} \nonumber
\\ &&=\exp \left\{ \frac i2\sum\limits_{a,b=1,..8}\int
d^4xd^4yJ^{\mu,a}\left(x\right) \widetilde{D}_{\mu \nu
}^{ab}(x-y)J^{\nu,b}\left(y\right) \right\}, \label{wick}
\end{eqnarray}
where the usual gluon propagator is denoted by $D_{\mu \nu
}^{ab}(x-y)$ and $\widetilde{D}_{\mu \nu }^{ab}(x-y)$ is the
modified one.

The result obtained for the modified propagator in the momentum
representation was,
\begin{equation}
\widetilde{D}_{\mu \nu }^{ab}\left(k\right) =\delta ^{ab}g_{\mu
\nu } \left[ \frac 1{k^2}-iC\delta \left(k\right) \right],
\label{prop1}
\end{equation}
In Eq. (\ref{prop1}), $C$ is the parameter associated with the
gluon condensate, and was determined to be real and non-negative
\cite{PRD,tesis}

The modification to the standard free ghost propagator, introduced
by the proposed initial state, was also calculated. Moreover,
after considering the free parameter in the proposed trial state
as real (\ref{Vacuum}), the ghost propagator was not modified.

\section{The Generating Functional and the Modified \\ Feynman
Rules, for $\alpha =1$}

In the present section the modified Feynman rules for the theory
are exposed. The generating functional of the Green functions, can
be written as

{\small
\begin{equation}
Z\left[ J,\bar{\eta},\eta,\bar{\xi},\xi \right] =\frac{\int
D\left(A,\bar{c },c,\bar{\Psi},\Psi \right) \exp \left\{ i\int
d^{4}x\left({\cal L}+J^{\mu,a}A_{\mu }^{a}+\overline{c}^{a}\eta
^{a}+\bar{\eta}^{a}c^{a}+{\bar{\Psi}} ^{i}{\xi
}^{i}+{\bar{\xi}}^{i}{\Psi }^{i}\right) \right\} }{\int D\left(A,
\bar{c},c,\bar{\Psi},\Psi \right) \exp \left\{ i\int d^{4}x{\cal
L}\right\} }, \label{z1}
\end{equation}}\noindent where sources have been introduced for all the fields in
the usual manner, and the effective Lagrangian is given by
\cite{Kugo,OjimaTex}
\begin{eqnarray}
{\cal L} &=&{\cal L}_{G}+{\cal L}_{Gh}+{\cal L}_{Q} \label{Lag} \\
{\cal L}_{G} &=&-\frac{1}{4}F_{\mu \nu }^{a}\left(x\right) F^{\mu
\nu,a}\left(x\right) -\frac{1}{2\alpha }\left(\partial ^{\mu
}A_{\mu }^{a}\left(x\right) \right) ^{2},\label{G} \\ {\cal
L}_{Gh} &=&-i\partial ^{\mu }\overline{c}^{a}\left(x\right) D_{\mu
}^{ab}\left(x\right) c^{b}\left(x\right),\label{FP} \\ {\cal
L}_{Q} &=&\bar{\Psi}^{i}\left(x\right) \left(i\gamma ^{\mu }D_{\mu
}^{ij}-m_{Q}\delta ^{ij}\right) \Psi ^{j}\left(x\right). \label{Q}
\end{eqnarray}

The sum over the six quark flavors will be omitted everywhere in
order to simplify the exposition, no confusion should arise for
it. The gluon field intensity has the usual form
\[
F_{\mu \nu }^{a}\left(x\right) =\partial _{\mu }A_{\nu
}^{a}\left(x\right) -\partial _{\nu }A_{\mu }^{a}\left(x\right)
+gf^{abc}A_{\mu }^{b}\left(x\right) A_{\nu }^{c}\left(x\right),
\]
$D_{\mu }^{ab},D_{\mu }^{ij}$ are the covariant derivatives in the
adjoint and fundamental representations respectively, for the
$SU(3)$ group
\begin{eqnarray*}
D_{\mu }^{ab}\left(x\right) &=&\partial _{\mu }\delta ^{ab}-
gf^{abc}A_{\mu }^{c}\left(x\right), \\ D_{\mu }^{ij}\left(x\right)
&=&\partial _{\mu }\delta ^{ij}-igT^{ij,a}A_{\mu }^{a}.
\end{eqnarray*}

It should be recalled that as argued in \cite{PRD,tesis} the
physical predictions for the value $\alpha =1$ of the gauge
parameter should have physical meaning whenever the adiabatic
connection of the interaction does not lead the evolving state out
the physical space. Therefore here this value of the parameter
will be fixed for this Chapter and the next one. The
generalization is advanced in Chapter 4.

Then in (\ref{z1}), the standard procedure of extracting out of
the functional integral the exponential of terms higher than
second order in the fields (vertexes) can be used. That is, thanks
to the recourse of exchanging the fields by functional derivatives
of the corresponding sources, and also calculating the remaining
Gaussian integral, the following expression for the generating
functional is obtained \cite{Faddeev}: {\small
\begin{eqnarray}
&&Z\left[ J,\bar{\eta},\eta,\bar{\xi},\xi \right] =N\exp \left\{
i\left[ \frac{S_{abc}^{G}}{3!i^{3}}\frac{\delta }{\delta
J_{a}}\frac{\delta }{\delta J_{b}}\frac{\delta }{\delta
J_{c}}+\frac{S_{abcd}^{G}}{4!i^{4}}\frac{\delta }{\delta
J_{a}}\frac{\delta }{\delta J_{b}}\frac{\delta }{\delta
J_{c}}\frac{ \delta }{\delta J_{d}}\right. \right. \label{funct2}
\\ &&\qquad \qquad \qquad\qquad \ \ \ \ \ +\left. \left.
\frac{S_{ras}^{Gh}}{2!i^{3}}\frac{\delta }{\delta
\bar{\eta}_{r}}\frac{\delta }{\delta J_{a}}\frac{\delta }{\delta
\left(-\eta _{s}\right) }+\frac{S_{iaj}^{Q}}{i^{3}}\frac{\delta
}{\delta \bar{\xi} _{i}}\frac{\delta }{\delta J_{a}}\frac{\delta
}{\delta \left(-\xi _{j}\right) }\right] \right\} Z_{0}\left[
J,\bar{\eta},\eta,\bar{\xi},\xi \right], \nonumber
\end{eqnarray}}\noindent with {\small
\begin{equation}
Z_{0}\left[ J,\bar{\eta},\eta,\bar{\xi},\xi \right] =\exp \left\{
i\frac{ J_{a}G_{G}^{ab}J_{b}}{2}+i\bar{\eta}_{r}G_{Gh}^{rs}\eta
_{s}+i\bar{\xi} _{i}G_{Q}^{ij}\xi _{j}\right\},
\end{equation}
and the normalization factor is {\small
\begin{eqnarray*}
&&N^{-1} =\left[ \exp \left\{ i\left[
\frac{S_{abc}^{G}}{3!i^{3}}\frac{ \delta }{\delta
J_{a}}\frac{\delta }{\delta J_{b}}\frac{\delta }{\delta J_{c}
}+\frac{S_{abcd}^{G}}{4!i^{4}}\frac{\delta }{\delta
J_{a}}\frac{\delta }{ \delta J_{b}}\frac{\delta }{\delta
J_{c}}\frac{\delta }{\delta J_{d}}\right. \right. \right.
\nonumber \\ &&\qquad \qquad \ \left. +\left. \left.
\frac{S_{ras}^{Gh}}{2!i^{3}}\frac{\delta }{ \delta
\bar{\eta}_{r}}\frac{\delta }{\delta J_{a}}\frac{\delta }{\delta
\left(-\eta _{s}\right) }+\frac{S_{iaj}^{Q}}{i^{3}}\frac{\delta
}{\delta \bar{\xi}_{i}}\frac{\delta }{\delta J_{a}}\frac{\delta
}{\delta \left(-\xi _{j}\right) }\right] \right\} Z_{0}\left[
J,\bar{\eta},\eta,\bar{\xi},\xi \right] \right]
_{J,\bar{\eta},\eta,\bar{\xi},\xi =0},\nonumber
\end{eqnarray*}}\noindent where use have been made of the DeWitt
compact notation \cite{Daemi}, in which a Latin letter
($a,b,...$), for a field symbolizes its space-time coordinates as
well as all its internal quantum numbers. The same index in a
source or a tensor symbolizes the same set of variables of the
kind of fields associated with this specific index. For example $
A^a= A^{a}_{\mu_a}(x_a),\ c^a=c^{a}(x_a)\ \text{and} \ \Psi^{i}=
\Psi^{i}(x_i)$. Such a procedure was useful for the calculations
and should not create confusion. As usual, repeated indexes
represent the corresponding space time integrals and contracted,
Lorentz, spinor or color components.

The following definitions have been used in (\ref{funct2}):

\begin{eqnarray*}
S_{ijk...}^{\alpha } &\equiv &\left(\frac{\delta }{\delta \Phi
_{i}}\frac{ \delta }{\delta \Phi _{j}}\frac{\delta }{\delta \Phi
_{k}}...S^{\alpha }\left(\Phi \right) \right) _{\Phi =0}\text{ \
for \ }\alpha = G,\ Gh,\ Q \text{ \ and }\Phi =A,\ \bar{c},\ c,\
\bar{\Psi}, \ \Psi \, \\ G_{\alpha }^{ij} &\equiv &-S_{\alpha
,ij}^{-1}\text{\qquad \qquad \qquad \qquad \qquad for \ } \alpha
=G,\ Gh,\ Q,
\end{eqnarray*}
where $G,\ Gh$ and $Q$ mean the gluon, ghost and quark parts of
the action respectively.

There is only one main element determining the difference of the
usual Perturbative QCD expansion and the modified one considered
in \cite{1995}-\cite{tesis}. It is related with the form of the
gluon propagator. As proposed in \cite{1995} there is an
additional term in the gluon propagator which is absent in the
standard expansion. In \cite{PRD,tesis}, such a term was shown to
be a consequence of a Wick expansion based in a state constructed
by acting on the usual vacuum with an exponential of zero momentum
pairs of gluon and ghost creation operators. By the side, the
ghost propagator \cite{PRD,tesis} could remain unmodified if the
parameter, free after fixing the form of the wave-function, is
taken as a real and positive one \cite{PRD,tesis}. Such a
selection will be employed here. In Sect. 3.2 possible
implications of the ghost propagator modification will be
discussed. The quark propagator is not affected in any way by the
gluon condensation as introduced in \cite{1995}-\cite{tesis}.

From a more technical point of view the operator quantization
employed in \cite{PRD,tesis} takes the ghost fields as satisfying
the conjugation properties defined by the Kugo and Ojima approach
\cite{Kugo,OjimaTex}. However, at the level of the Feynman diagram
expansion the difference with the standard procedure is only a
change of variables.

In accordance with above remarks and the results of
\cite{1995}-\cite{tesis} the diagram technique defining the
modified expansion has the following basic elements
\begin{center}
{\large Propagators}
\end{center}
\begin{eqnarray*}
G_{G}^{ab} &=&\delta ^{ab}g^{\mu _{a}\mu _{b}}\left[
\frac{1}{k^{2}+i\varepsilon} -iC\delta \left(k\right) \right], \\
G_{Gh}^{rs} &=&\delta ^{rs}\frac{\left(-i\right)
}{k^{2}+i\varepsilon}, \\ G_{Q}^{ij} &=&\delta ^{ij}\frac{m+p^{\mu
}\gamma _{\mu }}{\left(m^2-p^{2}-i\varepsilon\right) }.
\end{eqnarray*}

\newpage

\begin{center}
{\large vertexes}
\end{center}
\vspace{-0.2cm}
\begin{eqnarray*}
S_{abc}^{G} &=&\left(2\pi \right) ^{4}\delta
\left(k_{a}+k_{b}+k_{c}\right) \left(-i\right) gf^{abc}\left[
g_{\mu _{a}\mu _{c}}\left(k^{a}-k^{c}\right) _{\mu _{b}}+g_{\mu
_{a}\mu _{b}}\left(k^{b}-k^{a}\right) _{\mu _{c}}\right.
\\
&&\text{ \ \ \ \ \ \ \ \ \ \ \ \ \ \ \ \ \ \ \ \ \ \ \ \ \ \ \ \ \
\ \ \ \ \ \ \ \ \ \ \ \ \ \ \ }\left. +g_{\mu _{b}\mu _{c}}
\left(k^{c}-k^{b}\right) _{\mu _{a}}\right], \\ S_{abcd}^{G}
&=&\left(2\pi \right) ^{4}\delta
\left(k_{a}+k_{b}+k_{c}+k_{d}\right) \left(-g^{2}\right) \left[
f^{eab}f^{ecd}\left(g_{\mu _{a}\mu _{c}}g_{\mu _{b}\mu
_{d}}-g_{\mu _{b}\mu _{c}}g_{\mu _{a}\mu _{d}}\right) +\right. \\
&&\text{ \ \ \ \ \ \ \ \ \ \ \ \ \ \ \ \ \ \ \ \ \ \ \ \ \ \ \ \ \
\ \ \ \ \ \ \ \ \ \ \ \ \ \ }+f^{ecb}f^{ead}\left(g_{\mu _{a}\mu
_{c}}g_{\mu _{b}\mu _{d}}-g_{\mu _{d}\mu _{c}}g_{\mu _{a}\mu
_{b}}\right) \\ &&\text{ \ \ \ \ \ \ \ \ \ \ \ \ \ \ \ \ \ \ \ \ \
\ \ \ \ \ \ \ \ \ \ \ \ \ \ \ \ \ \ \ \ \ \ }\left.
+f^{edb}f^{eca}\left(g_{\mu _{d}\mu _{c}}g_{\mu _{a}\mu
_{b}}-g_{\mu _{b}\mu _{c}}g_{\mu _{a}\mu _{d}}\right) \right], \\
S_{ras}^{Gh} &=&\left(2\pi \right) ^{4}\delta
\left(k_{r}+k_{a}+k_{s}\right) gf^{ras}\left(k^{r}-k^{s}\right)
_{\mu _{a}}, \\ S_{iaj}^{Q} &=&\left(2\pi \right) ^{4}\delta
\left(k_{i}+k_{a}+k_{j}\right) gT_{ij}^{a}\gamma _{\mu _{a}}.
\end{eqnarray*}

The propagators and vertexes will be represented by,

\begin{figure}[h]
\begin{center}
\includegraphics[scale=0.37,angle=0]{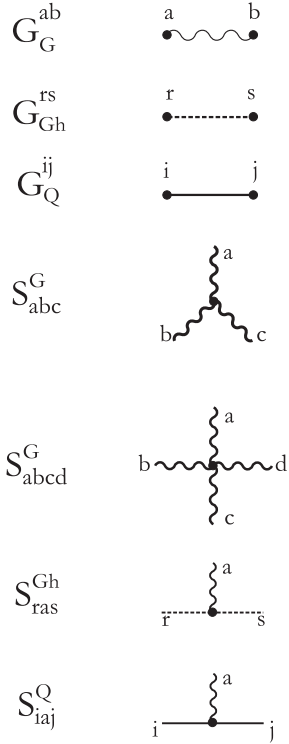}
\end{center}
\vspace{-0.5cm} \caption{Propagator and vertex representations.}
\label{Diagram}
\end{figure}

Being defined the Feynman rules under consideration, in the next
section they will be applied to the calculation of various
quantities of physical interest.

\chapter{Calculations within the Modified Perturbation Theory}

Some magnitudes of physical interest are calculated within the
modified perturbation theory, up to the $g^2$ order, and in the
approximation described in the Introduction. That is disregarding
all the terms not including the gluon condensate. The mean value
of $G^2$ (Section 3.1) and the gluon mass (Section 3.2) are
determined improving previous calculations \cite{1995}. In Section
3.3 the constituent quarks masses are calculated. Finally the
effective potential \cite{1995} is obtained in terms of the mean
value of $G^{2}$ (Section 3.4).

\section{Mean value of $G^{2}$}

The calculation of the mean value of $G^2$ is considered in this
section. It completes the former evaluation done in \cite{1995} in
the tree approximation. The result will be employed afterwards for
fixing the value of the condensation parameter $g^2 C$ in order to
obtain the currently accepted value for $\langle g^2 G^2\rangle$.
As it was stated in the introduction, the quantities will be
exactly calculated disregarding all the terms not containing the
influence of the condensate. It can be remarked that all the terms
up to the $g^2$ order considered here, in which the condensate
parameter enters and using dimensional regularization, do not need
for renormalization. This result is another resemblance with a
quasi-classical approximation.

For the calculation of $\langle G^2\rangle$ the following
expression will be used

{\small
\begin{equation}
\left\langle 0\right| S_{g}\left| 0\right\rangle =\frac{\int
D\left(A,\bar{c },c,\bar{\Psi},\Psi \right) S_{g}\left[ A\right]
\exp \left\{ i\int d^{4}x\left({\cal L}+J^{\mu,a}A_{\mu
}^{a}+\overline{c}^{a}\eta ^{a}+\bar{
\eta}^{a}c^{a}+{\bar{\Psi}}^{i}{\xi }^{i}+{\bar{\xi}}^{i}{\Psi
}^{i}\right) \right\} }{\int D\left(A,\bar{c},c,\bar{\Psi},\Psi
\right) \exp \left\{ i\int d^{4}x{\cal L}\right\} }, \label{sg}
\end{equation}}\noindent where $S_{g}\left[ A\right]$ represents
the gluon part of the action in absence of the gauge breaking term
depending on the gauge parameter $\alpha$, that is

\begin{equation}
S_{g}\left[ A\right] =-\frac{1}{4}\int d^{4}xF_{\mu \nu
}^{a}\left(x\right) F^{\mu \nu,a}\left(x\right).
\end{equation}

Through the use again of the trick of changing the fields inside
the functional integral by derivatives in corresponding sources,
it follows for (\ref{sg})

{\small
\begin{equation}
\left\langle 0\right| S_{g}\left| 0\right\rangle =\left\{ \left[
\frac{ S_{ab}^{g}}{2i^{2}}\frac{\delta }{\delta J_{a}}\frac{\delta
}{\delta J_{b}}+ \frac{S_{abc}^{G}}{3!i^{3}}\frac{\delta }{\delta
J_{a}}\frac{\delta }{\delta J_{b}}\frac{\delta }{\delta
J_{c}}+\frac{S_{abcd}^{G}}{4!i^{4}}\frac{\delta }{\delta
J_{a}}\frac{\delta }{\delta J_{b}}\frac{\delta }{\delta
J_{c}}\frac{ \delta }{\delta J_{d}}\right] Z\left[
J,\bar{\eta},\eta,\bar{\xi},\xi \right] \right\}
_{J,\bar{\eta},\eta,\bar{\xi},\xi =0} \label{des1}
\end{equation}}\noindent where $Z\left[ J,\bar{\eta},\eta,\bar{\xi},\xi \right]$
is given by the expression (\ref{funct2}). In the momentum
representation $S_{ab}^{g}$ has the form

\[
S_{ab}^{g}=\left(2\pi \right) ^{4}\delta \left(k_{a}+k_{b}\right)
\delta ^{ab}\left[ -\left(g_{\mu _{a}\mu _{b}}k_{a}^{2}-k_{a,\mu
_{a}}k_{a,\mu _{a}}\right) \right],
\]
and it will be represented in the calculations by

\begin{figure}[h]
\begin{center}
\includegraphics[scale=0.07,angle=0]{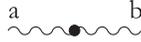}
\end{center}
\vspace{-0.7cm} \caption{Representation of $S_{ab}$.}
\end{figure}

It should be noticed that the contribution of diagrams including
any kind of particles (which involve the condensate effects), to
the vacuum expectation value of $S_{g}$, needs for the evaluation
of loop integrals which only can be properly accounted after
renormalization. However, in the present situation for these
contributions to $\left\langle 0\right| S_{g}\left|
0\right\rangle$, it happens that the loop integrals which appears
are evaluated at zero momentum. Then, the results vanish using the
dimensional regularization and thus do not depend on the
renormalization conditions.

After computing all the diagrams appearing in the $\left\langle
0\right| S_{g}\left| 0\right\rangle$ expression, with the remarks
made above, was obtained that the only diagram contributing is the
one appearing in the Figure \ref{g2}. That is, all the diagrams
not considered in earlier works have a null contribution. The
details of the calculation are exposed in the Appendix A.
\begin{figure}[h]
\begin{center}
\includegraphics[scale=0.24,angle=0]{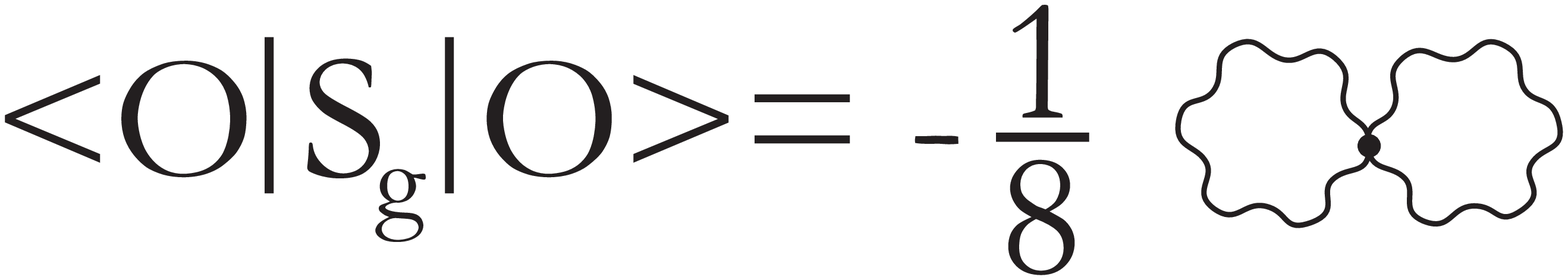}
\end{center}
\vspace{-0.7cm} \caption{Diagram contributing to $\left\langle
0\right| S_{g}\left| 0\right\rangle$.} \label{G2}
\end{figure}

The diagram in Figure \ref{G2} was calculated before
\cite{1995,PRD}, the result is

\begin{equation}
\left\langle 0\right| S_{g}\left| 0\right\rangle =-\frac{1}{4}\int
d^{4}x \frac{288g^{2}C^{2}}{\left(2\pi \right) ^{8}},
\end{equation}
which, in turns implies for the mean value of $G^{2}$ the result

\begin{equation}
\langle G^{2}\rangle =\frac{288g^{2}C^{2}}{\left(2\pi \right)
^{8}}. \label{g2m}
\end{equation}

\noindent Finally, making use of a currently estimated value for
$\langle g^{2}G^{2}\rangle$:

\[
\langle g^{2}G^{2}\rangle \cong 0.5\left(GeV/c^{2}\right) ^{4},
\]
the parameter $g^{2}C$ is determined to have the value:

\begin{equation}
g^{2}C=64.94\ \left(GeV/c^{2}\right) ^{2}. \label{g2}
\end{equation}

The fixing of the condensation parameter allows to investigate the
physical predictions of the modified expansion in the above
described approximation which does not require to introduce the
renormalization.

\section{Gluon Mass}

In the present section it is exposed the calculation for the gluon
mass.

The expression for the effective action of the gluon sector, in
the one loop approximation or up to $g^2$ order \cite{Daemi}, is

\begin{equation}
\Gamma^{G} \left[\phi _{c}\right] =S_{G}\left[\phi _{c}\right]
+\frac{1}{2i}\ln \det G_{G}\left[\phi _{c}\right], \label{effect}
\end{equation}
in Feynman diagrams this expression can be written in the form,

\begin{figure}[h]
\begin{center}
\includegraphics[scale=0.18,angle=0]{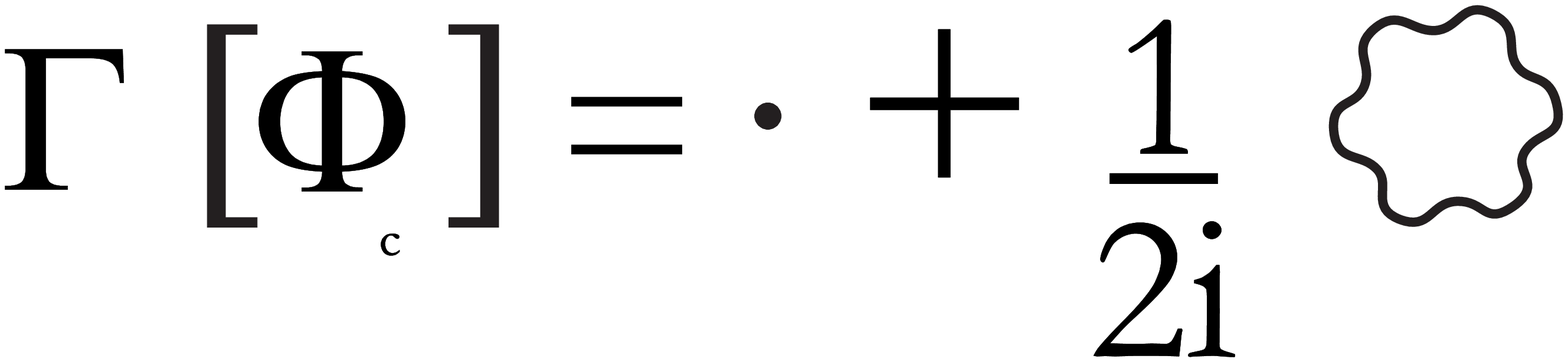}
\end{center}
\vspace{-0.7cm} \caption{The effective action.}
\end{figure}

The second functional derivative of (\ref{effect}), in the gluon
fields, defines the two point proper Green function or the inverse
of the gluon propagator. It can be written in the form

\begin{equation}
\Gamma _{,ij}^{G}\left[\phi _{c}\right] =S_{,ij}^{G}\left[\phi
_{c}\right] +\frac{ 1 }{2i}\left(S_{,ipqj}^{G}\left[\phi
_{c}\right] G_{G}^{pq}+S_{,ipq}^{G}\left[\phi _{c}\right]
G_{G}^{qr}G_{G}^{ps}S_{,srj}^{G}\left[\phi _{c}\right] \right).
\label{exp1}
\end{equation}
where the quark and ghost fields do not contribute in the
approximation considered. That is, their diagrams do not involve
the gluon condensate.

As it was stated in a previous work \cite{1995}, the interest is
centered in the case of zero mean values of the gluonic field (as
it is required by the Lorentz invariance), so the expression
(\ref{exp1}) takes the form,

\begin{equation}
\Gamma _{,ij}^{G}\left[0\right] =S_{ij}^{G}+\frac{1}{2i}\left(
S_{ipqj}^{G}G_{G}^{pq}+S_{ipq}^{G}G_{G}^{qr}G_{G}^{ps}S_{srj}^{G}\right),
\end{equation}

\noindent or diagrammatically

\begin{figure}[h]
\begin{center}
\includegraphics[scale=0.43,angle=0]{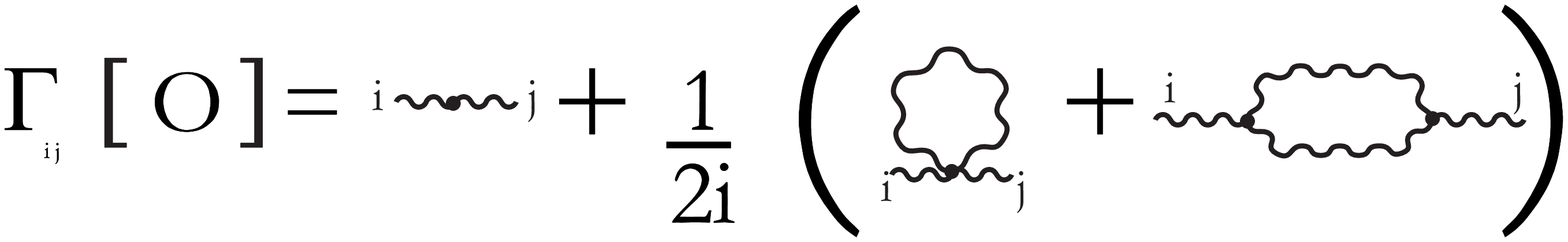}
\end{center}
\caption{Two point proper Green function.}
\end{figure}

Writing the expressions in the momentum representation, the
following results are obtained for the gluon condensate
contributions

{\setlength\arraycolsep{0.9pt}
\begin{eqnarray}
S_{ij}^{G} &=&-\delta ^{a_{i}a_{j}}g_{\mu _{i}\mu _{j}}p^{2}, \\
S_{ipqj}^{G}G_{G}^{pq} &=&i\frac{6g^{2}CN}{\left(2\pi \right)
^{4}}\delta ^{a_{i}a_{j}}g_{\mu _{i}\mu _{j}}, \\
S_{ipq}^{G}G_{G}^{qr}G_{G}^{ps}S_{srj}^{G}
&=&i\frac{g^{2}CN}{\left(2\pi \right) ^{4}}\delta
^{a_{i}a_{j}}\left(-10g_{\mu _{i}\mu _{j}}+4\frac{p_{\mu
_{i}}p_{\mu _{j}}}{p^{2}}\right).
\end{eqnarray}
}

The sum of the three contributions is

\begin{equation}
\Gamma _{,ij}^{G}\left[0;p\right] =\delta
^{a_{i}a_{j}}\left[-\left(p^{2}+\frac{2g^{2}CN}{\left(2\pi \right)
^{4}}\right)\left(g_{\mu _{i}\mu _{j}}-\frac{p_{\mu _{i}}p_{\mu
_{j}}}{p^{2}}\right)-p^{2}\frac{p_{\mu _{i}}p_{\mu _{j}}}{p^{2}}
\right]. \label{exp2}
\end{equation}

For calculating the previous expressions the following relation,
between the structure constants for a gauge group
$SU\left(N\right)$, was employed
\begin{equation}
f^{a_{i}a_{l}a_{k}}f^{a_{j}a_{l}a_{k}}=N\delta ^{a_{i}a_{j}}.
\end{equation}

Considering the specific group $SU\left(3\right)$, the expression
(\ref{exp2}) takes the form

\begin{equation}
\Gamma _{,ij}^{G}\left[0;p\right] =\delta
^{a_{i}a_{j}}\left[-\left(p^{2}+\frac{6g^{2}C}{\left(2\pi \right)
^{4}}\right)\left(g_{\mu _{i}\mu _{j}}-\frac{p_{\mu _{i}}p_{\mu
_{j}}}{p^{2}}\right)-p^{2}\frac{p_{\mu _{i}}p_{\mu _{j}}}{p^{2}}
\right].\label{dis1}
\end{equation}

The eigenvalues of the previous expression define the dispersion
relations, and have the form

\begin{equation}
p^{2}=0,\ \ \ p^{2}-m^{2}_{G}=0,
\end{equation}
where

\begin{equation}
m_{G}^{2}=-\frac{6g^{2}C}{\left(2\pi \right)
^{4}}=-0.25\left(GeV/c^{2} \right)^{2}. \label{m2}
\end{equation}

Therefore, as the parameter $C$ was defined before as real and
positive, in the framework of the initial state considered
\cite{PRD,tesis}, it follows that the transverse gluon mass
correction becomes tachyonic. The ability of a tachyonic mass in
producing improvements in models for the inter quark potential has
been recently argued in the literature \cite{tachyon1,tachyon2}.
The effect is related with the introduction of a linearly rising
term as a first correction to the Coulomb potential in the
massless case \cite{tachyon2}. Therefore, the tachyonic result
arising here for the gluon mass, appears to be of interest for an
attempt of deriving from the present framework the already
existing successful phenomenological bound state models for mesons
\cite{sommer1}-\cite{El-Hady}.

Finally it should be mentioned that if a modification of the ghost
propagator is considered, then appears a longitudinal mass term
for the propagator, as consequence of the diagram in Fig.
\ref{long} and the polarization operator (gluon self-energy)
becomes non-transverse. This result means a violation of a Ward
identity representing the gauge invariance, by the presence of the
ghost condensate. Thus, it follows from the present results that
the selection of the parameter for the initial state in
\cite{PRD,tesis} was the appropriate one for the fulfillment of a
gauge invariance condition at this stage. The physical meaning of
alternative procedures for constructing the initial state need
however a clarification.

\begin{figure}[h]
\begin{center}
\includegraphics[scale=0.13,angle=0]{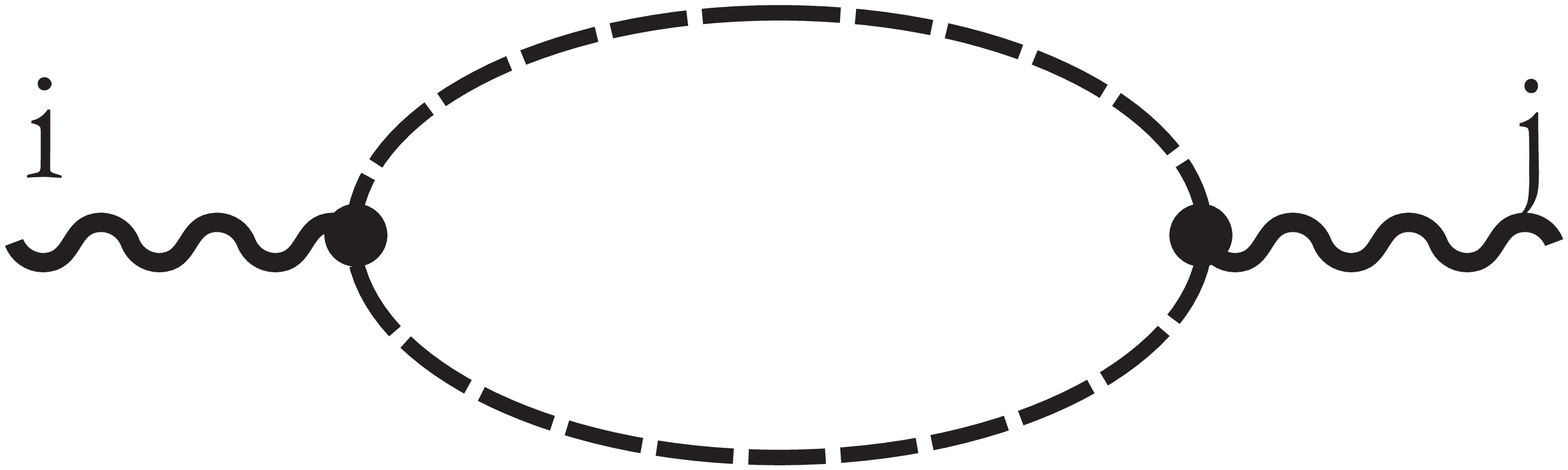}
\caption{Ghost contribution to the gluon self-energy.}
\label{long}
\end{center}
\end{figure}

\section{Quark Masses}

The effects produced by the condensate, on the quark masses, are
considered in this section. The full inverse propagator for
quarks, in terms of the corresponding mass operator, can be
expressed as \cite{Muta}

\begin{equation}
G_{Q}^{-1}\left(p\right) =\left(m_{Q}-p^{\mu }\gamma _{\mu
}-\Sigma \left(p\right) \right),
\end{equation}

\noindent where the color index identity matrix has been omitted
and $\Sigma\left(p\right)$ is the mass operator. Its expression is
represented by the diagram in Figure \ref{autoq}.

\begin{figure}[h]
\begin{center}
\includegraphics[scale=0.20,angle=0]{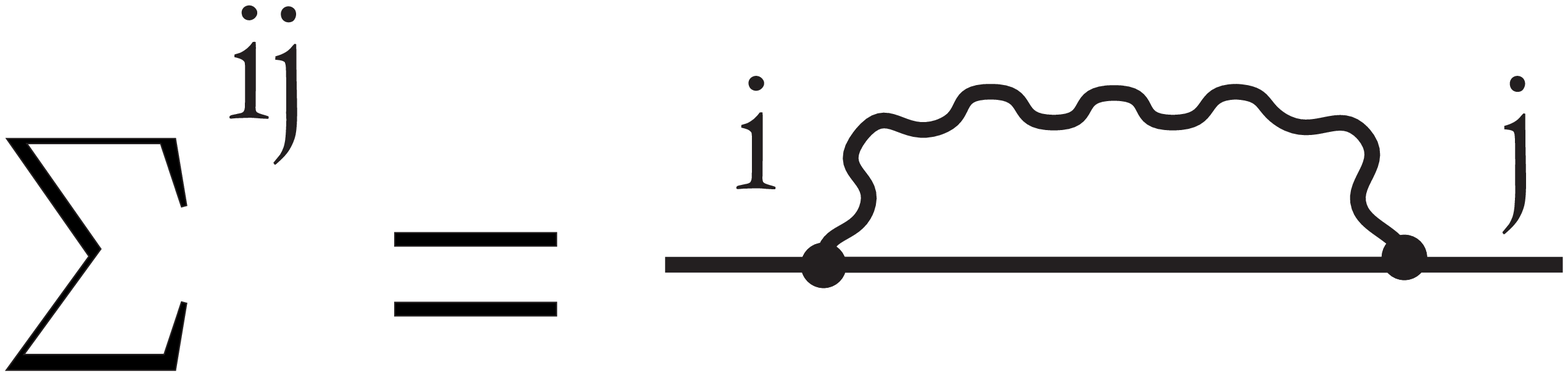}
\end{center}
\caption{Diagram for $\Sigma\left(p\right)$.} \label{autoq}
\end{figure}

The exact form of $\Sigma \left(p\right)$ up to the $g^2$ order,
considering the modified Feynman rules, is

\begin{equation}
\Sigma \left(p\right) =g^{2}C_{F}\int \frac{d^{4}k}{\left(2\pi
\right) ^{4}i} \frac{\gamma _{\mu }\left(m_{Q}+\left(p-k\right)
^{\alpha }\gamma _{\alpha }\right) \gamma _{\nu }G_{G}^{\mu \nu
}\left(k\right) }{\left(m_{Q}^{2}- \left(p-k\right) ^{2}\right) },
\label{auto}
\end{equation}
in which the algebra for the color indexes has been calculated
\cite{Muta}, leading to the following expression for the constant
$C_{F}$ ($SU\left(N\right)$)

\[
C_{F}=\frac{N^{2}-1}{2N},
\]
for the case of interest $SU\left(3\right)$, it reduces to
$C_{F}=\frac{4}{3}$.

Taking into account the modified gluon propagator and the standard
relations for the $\gamma$ matrices (which here are adopted the
same as in Ref. \cite{Muta}) the full one loop mass operator
expression can be simplified to
\begin{equation}
\Sigma \left(p\right) =g^{2}C_{F}\int \frac{d^{4}k}{\left(2\pi
\right) ^{4}i} \frac{2\left(2m_{Q}-\left(p-k\right) ^{\alpha
}\gamma _{\alpha }\right) }{ \left(m^{2}_{Q}-\left(p-k\right)
^{2}\right) }\left(\frac{1}{k^{2}} -iC\delta \left(k\right)
\right).
\end{equation}

Here, as before, the term not involving the condensate won't be
considered. In this case the expression for $\Sigma
\left(p\right)$ reduces to

\begin{equation}
\Sigma \left(p\right) =-\frac{g^{2}C_{F}C}{\left(2\pi \right)
^{4}}\frac{ 2\left(2m_{Q}-p^{\alpha }\gamma _{\alpha }\right)
}{\left(m^{2}_{Q}-p^{2} \right) }.
\end{equation}

Then the expression for the inverse quark propagator reduces to

\begin{equation}
G_{Q}^{-1}\left(p\right) =m_{Q}\left(1+2\frac{
M^{2}}{\left(m_{Q}^{2}-p^{2}\right) }\right) -p^{\mu }\gamma _{\mu
}\left(1+ \frac{M^{2}}{\left(m_{Q}^{2}-p^{2}\right) }\right),
\label{prop2}
\end{equation}
where it is defined

\[
M^{2}=\frac{2g^{2}C_{F}C}{\left(2\pi \right) ^{4}}=0.11\left(
GeV/c^{2}\right) ^{2},
\]
in which the numerical value has been obtained considering
(\ref{g2}).

The zeros of the determinant associated to the inverse propagator
(\ref{prop2}) allow to determine the effects of the condensate
(reflected by the parameter $M$), on the effective mass or mass
shell of quarks. Now is useful to introduce an index $s, s=1...6$
for each kind of quark characterized by its particular current
mass $m_{Q_s}$.

The mass shell obtained in the present case is

\begin{equation}
p^{2}-m_{q_{s,i}}^{2}=0,
\end{equation} for $m_{q_{s,i}}^{2}$ are
obtained three different analytical expressions arising from the
solutions of a cubic equation. They are

\begin{eqnarray*}
m_{q_{s,1}}^{2} &=&A^{\frac{1}{3}}-B+m^{2}_{Q_s}+\frac{2}{3}M^{2},
\\ m_{q_{s,2}}^{2}
&=&-\frac{1}{2}A^{\frac{1}{3}}+\frac{1}{2}B+m^{2}_{Q_s}+\frac{2}{
3} M^{2}+\frac{1}{2}i\sqrt{3}\left(A^{\frac{1}{3}}+B\right), \\
m_{q_{s,3}}^{2}
&=&-\frac{1}{2}A^{\frac{1}{3}}+\frac{1}{2}B+m^{2}_{Q_s}+\frac{2}{
3} M^{2}-\frac{1}{2}i\sqrt{3}\left(A^{\frac{1}{3}}+B\right),
\end{eqnarray*}
in which $A$, $B$ are given by the following expressions

\begin{eqnarray*}
A&=&\frac{5}{6}m^{2}_{Q_s}M^{4}-\frac{1}{27}M^{6}+\frac{1}{18}\sqrt{
96m^{6}_{Q_s}M^{6}+177m^{4}_{Q_s}M^{8}-12m^{2}_{Q_s}M^{10}}, \\ B
&=&\frac{\frac{2}{3}m^{2}_{Q_s}M^{2}-\frac{1}{9}M^{4}}{A^{\frac{1}{3}}}.
\end{eqnarray*}

In order to find the dependence of the effective quark masses
$m_{q_{s,i}}^{2}$ as functions of the current mass parameters
$m_{Q_s}$, an analysis for the various dispersion relations that
appeared, was done. It follows that there is only one solution
having a squared mass being positive for arbitrary values of the
current quark masses. The existence of this real and positive
solution for the squared mass is possible because of the real and
positive character of $C$, as was proved previously
\cite{PRD,tesis} and fixed through $\langle g^2G^2\rangle$. The
other solutions for the masses become complex for certain values
of the current masses $m_{Q_s}$, and there is no a clear
understanding of their meanings.

For the purely real solution the value of the effective quark
masses $m_{q_s}=f\left(m_{Q_s}\right)$ as functions of $m_{Q_s}$
is shown in the Figure \ref{Mass1}. The graph is plotted for the
region $m_{Q_s}<2\ GeV/c^{2}$ which contains the current mass
values of the $u,\ d,\ s$ and $c$ quarks.

\begin{figure}[h]
\begin{center}
\includegraphics[scale=0.4,angle=-90]{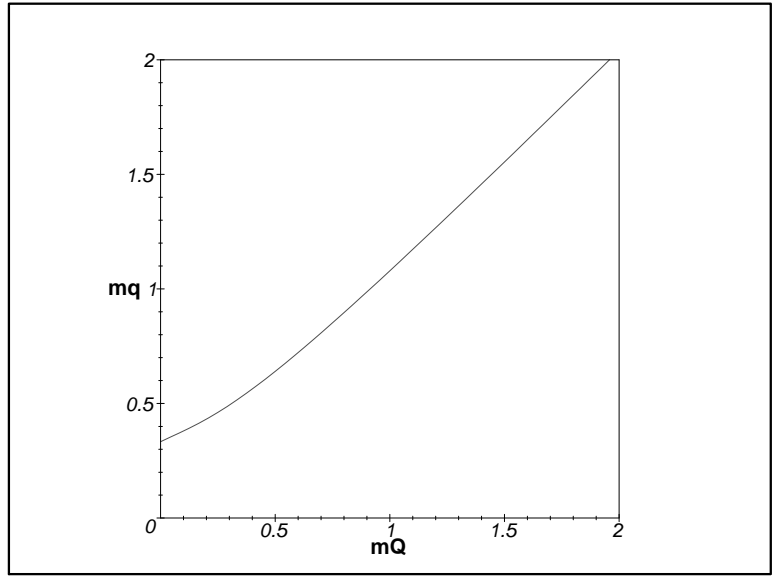}
\end{center}
\vspace{-0.7cm} \caption{Real solution for the quark mass as a
function of the Lagrangian mass (masses in $GeV/c^2$).}
\label{Mass1}
\end{figure}

As it can be appreciated in the picture, the calculated effective
quark masses for light flavors ($u,\ d$ and $s$) are clearly
predicting the values of the quark masses being in use in the
constituent quark models of hadrons. That is, the light quarks get
a weight of near one third of the nucleon mass. In table
\ref{table1} the mass values obtained, for each quark flavor, are
shown. The reported lower and upper values for the current masses
are denoted by $m_{Low}^{Exp}$ and $m_{Up}^{Exp}$, respectively
\cite{Report}. The correspondingly calculated constituent masses
are indicated by $m_{Low}^{Th}$ and $m_{Up}^{Th}$; finally $
m_{Med}^{Th}$ is the evaluated mean value of the constituent
masses.

\begin{table}[h]
\caption{Quark mass values ($MeV/c^2 $) in presence of the
condensate.} \label{table1}
\begin{center}
\begin{tabular}{||c||c||c||c||c||c||}
\hline\hline $Quarks$ & $m_{Low}^{Exp}$ & $m_{Up}^{Exp}$ &
$m_{Low}^{Th}$ & $m_{Up}^{Th}$ & $m_{Med}^{Th}$ \\ \hline\hline Up
(u) & 1.5 & 5 & 334.1 & 335.8 & 334.9 \\ \hline\hline Down (d) & 3
& 9 & 334.8 & 337.8 & 336.3 \\ \hline\hline Strange (s) & 60 & 170
& 361.3 & 415.1 & 388.2 \\ \hline\hline Charmed (c) & 1100 & 1400
& 1172.6 & 1457.9 & 1315.2 \\ \hline\hline Bottom (b) & 4100 &
4400 & 4120.3 & 4418.9 & 4269.6 \\ \hline\hline Top (t) & 168600 &
179000 & 168600.5 & 179000.5 & 173800.5 \\ \hline\hline
\end{tabular}
\end{center}
\end{table}

From the global properties of table \ref{table1} it can be
observed that the main effect of the gluon condensate seems to be
dressing the light quarks with a cloud of gluons having a total
mass of one third of the nucleon mass. Then, the results point in
the direction of the mainly glue nature of the constituent quark
masses of the $u,\ d$ and $s$ quarks within many baryon resonance.
These results support the idea that a modified perturbative
expansion like the one being considered, in which the effect of
the gluon condensate has been incorporated, would be able to
predict with reasonably good approximation properties of the low
energy strong interactions. Similar values for the constituent
quarks masses have been obtained by a different method in
\cite{Steele}.

In tables \ref{table2} and \ref{table3} are shown the masses for
baryon and meson ground states \cite{Report}, compared with the
values obtained for them by a theoretical estimate obtained as the
simple addition of the calculated constituent masses for each of
their known constituent quarks.

\begin{table}[tbp]
\caption{Experimental and theoretical masses ($ MeV/c^2$) for
baryonic resonances in their ground states.} \label{table2}
\begin{center}
\begin{tabular}{||r||r||r||r||}
\hline\hline Baryon & Exp.Val. & Th.Val. & Rel.Err. \\
\hline\hline p(uud) & 938.27231 & 1006.2 & 7.2 \\ \hline\hline
n(udd)  & 939.56563 & 1007.5 & 7.2 \\ \hline\hline $\Lambda $(uds)
 & 1115.683 & 1059.4 & 5.0 \\ \hline\hline $\Sigma ^{+}$(uus)
 & 1189.37 & 1058.1 & 11.0 \\ \hline\hline $\Sigma ^{0}$(uds)
 & 1192.642 & 1059.4 & 11.2 \\ \hline\hline $\Sigma ^{-}$(dds)
 & 1197.449 & 1060.8 & 11.4 \\ \hline\hline $\Xi ^{0}$(uss)
 & 1314.9 & 1111.4 & 15.5 \\ \hline\hline $\Xi ^{-}$(dss)
 & 1321.32 & 1112.7 & 15.8 \\ \hline\hline $\Omega ^{-}$(sss)
 & 1642.45 & 1164.6 & 29.1 \\ \hline\hline $\Lambda _{c}^{+}$(udc)
 & 2284.9 & 1986.5 & 13.1 \\ \hline\hline $\Xi _{c}^{+}$(usc)
 & 2465.6 & 2038.4 & 17.3 \\ \hline\hline $\Xi _{c}^{0}$(dsc)
 & 2470.3 & 2039.7 & 17.4 \\ \hline\hline $\Omega _{c}^{0}$(ssc)
 & 2704 & 2091.6 & 22.6 \\ \hline\hline $\Lambda _{b}^{0}$(udb)
 & 5624 & 4940.8 & 12.2 \\ \hline\hline
\end{tabular}
\end{center}
\end{table}

\begin{table}[tbp]
\caption{Experimental and theoretical masses ($MeV/c^2$) for
vector messon resonances in their ground states.} \label{table3}
\begin{center}
\begin{tabular}{||r||r||r||r||}
\hline\hline Meson \ \ \ & Exp.Val. & Th.Val. & Rel.Err. \\
\hline\hline $\rho
\left(\frac{u\overline{u}-d\overline{d}}{\sqrt{2}}\right) $ &
770.0 & 671.2 & 12.8 \\ \hline\hline $\varpi \left(
\frac{u\overline{u}+d\overline{d}}{\sqrt{2}}\right) $ & 781.94 &
671.2 & 14.2 \\ \hline\hline $\phi \left(s\overline{s}\right) $ \
\ \ \ & 1019.413 & 776.4 & 23.8 \\ \hline\hline $J/\psi \left(
1S\right) \left(c\overline{c}\right) $ & 3096.88 & 2630.5 & 15.1
\\ \hline\hline $Y\left(1S\right) \left(b\overline{b}\right) $ &
9460.37 & 8539.1 & 9.7 \\ \hline\hline
\end{tabular}
\end{center}
\end{table}

\newpage

As it can be appreciated the results obtained for the vector meson
and baryon resonances are only estimates. However, their values
are in good correspondence with the reported experimental values.
This situation is more curious taking into account: 1) The great
errors in the quark current masses \cite{Report} 2) The standard
one loop self energy contribution to the propagator was not
included. The major differences between the experimental values
and the theoretical calculations are obtained for resonances
containing the strange quark. The improvement of these results
could be introduced from higher order corrections, because the
strange quark is located between the almost massless up, down and
the heavy quarks, then the one loop correction could not be enough
to obtain the total influence of the condensate on that quark. It
can be noticed that the validity of the results obtained implies
that the binding energy contribution to the baryon rest masses is
small. This property then could justify a no relativistic
description of such resonances.

\section{Effective Potential}

In the present section is calculated the effective potential as a
function of the mean value of $G^{2}$. The approximation
considered is the same advanced in a previous work \cite{1995},
and the result is improved through the relation between the mean
value of $G^2$ and the gluon mass obtained in previous sections.
It is shown that the condensate is spontaneously generate from the
standard vacuum, as in the Savvidy models \cite{Savvidy1}.

The expression for the effective potential, considering the
presence of a mean field, was obtained by R. Jackiw \cite{Jackiw}
and has the form
\begin{equation}
V\left(\phi _{c}\right) =V_{0}\left(\phi _{c}\right)
+\frac{1}{2i}\int \frac{d^{4}k}{\left(2\pi \right) ^{4}}\ln \det
\left(-S_{,ij}\left[ \phi _{c};k\right] \right) +i\left\langle
\exp \left(i\int d^{4}x \mathcal{L}_{i}\left[ \phi _{c},\phi
\right] \right) \right\rangle, \label{V1}
\end{equation}
as it was mentioned before, in the present work is analyzed the
case for null mean values of gluonic fields ($\phi_{c}=0$), as it
is required by the Lorentz invariance. The first term in the
previous expression is $\langle G^{2}\rangle / 4$, the second term
represents the quantum corrections to the effective potential up
to one loop approximation, and the last term represents the higher
order corrections.

The approximation considered in the present section \cite{1995},
is the one in which all the mass insertions, consequence of the
gluon condensate, are introduced in the one loop term of the
expression (\ref{V1}). That is, it is substituted the
$S_{,ij}^{G}\left[ 0;k\right]$ by the $\Gamma_{,ij}^{G}\left[
0;k\right]$ calculated in Section 3.2, and it is like considering
$\Gamma_{,ij}^{G}\left[ 0;k\right]$ as a modified tree
approximation for the second derivative of the action under the
influence of the condensate. These corrections appear in an
exact calculation from the expansion of the third term of
(\ref{V1}), that won't be taken into account in what follows. In
this case the expression (\ref{V1}) takes the form
\begin{equation}
V\left(\langle G^{2}\rangle \right) = \frac{\langle
G^{2}\rangle}{4} + \frac{1}{2i}\int \frac{d^{4}k}{\left(2\pi
\right) ^{4}}\ln \det \left(-\Gamma _{ij}^{G}\left[ 0;k\right]
\right),
\end{equation}
that considering the relation
\begin{equation}
\ln \det \left(-\Gamma _{ij}^{G}\left[ 0;k\right] \right)
=\text{Tr}\ln \left(-\Gamma _{ij}^{G}\left[ 0;k\right] \right),
\end{equation}
can be written as
\begin{equation}
V\left(\langle G^{2}\rangle \right) = \frac{\langle
G^{2}\rangle}{4} +\frac{1}{2i}\int \frac{d^{4}k}{\left(2\pi
\right) ^{4}}\text{Tr}\ln \left(-\Gamma _{ij}^{G}\left[ 0;k\right]
\right),
\end{equation}

In order to calculate the previous integral, the relation
(\ref{dis1}) can be written in the form
\begin{equation}
\Gamma _{,ij}^{G}\left[ 0;k\right] =-\delta ^{a_{i}a_{j}}\left[
\left(k^{2}-m_{G}^{2}\right) P_{T}+k^{2}P_{L}\right],
\end{equation}
where $P_{T}$ is the transverse projection operator
\begin{equation}
P_{T}=g_{\mu _{i}\mu _{j}}-\frac{k_{\mu _{i}}k_{\mu _{j}}}{k^{2}},
\end{equation}
and $P_{L}$ is the longitudinal projection operator
\begin{equation}
P_{L}=\frac{k_{\mu _{i}}k_{\mu _{j}}}{k^{2}}.
\end{equation}

These operators satisfy the following relations:
\begin{eqnarray}
\left(P_{T}\right) ^{N} &=&P_{T},\text{ \ for }N>0 \\ \left(
P_{L}\right) ^{N} &=&P_{L},\text{ \ for }N>0 \\ P_{T}\cdot P_{L}
&=&0, \\ \text{Tr}(\delta ^{a_{i}a_{j}}P_{T}) &=&8\times 3=24, \\
\text{Tr}(\delta ^{a_{i}a_{j}}P_{L}) &=&8\times 1=8.
\end{eqnarray}

Then it is possible to write
\begin{eqnarray}
\text{Tr}\ln \left(-\Gamma _{ij}^{G}\left[0;k\right] \right)
&=&\text{Tr}\ln \left(\delta ^{a_{i}a_{j}}\left[\left(
k^{2}-m^{2}_{G}\right) P_{T}+k^{2}P_{L} \right] \right)\nonumber
\\ &=&24\ln \left(k^{2}-m^{2}_{G}\right) +8\ln \left(k^{2}\right),
\end{eqnarray}
and considering that
\begin{equation}
\int \frac{d^{4}k}{\left(2\pi \right) ^{4}i}\ln \left(
k^{2}\right) =const,
\end{equation}
the attention will be centered in the term
\begin{equation}
\int \frac{d^{4}k}{\left(2\pi \right) ^{4}i}\ln \left(
k^{2}-m^{2}_{G}\right) =\int \frac{d^{4}k}{\left(2\pi \right)
^{4}i}\ln \left(1-\frac{m^{2}_{G}}{ k^{2} }\right) +const.
\end{equation}

For the calculation is defined:
\begin{equation}
T_{m}=\int \frac{d^{4}k}{\left(2\pi \right) ^{4}i}\ln \left(
1-\frac{ m^{2}_{G}}{ k^{2}}\right). \label{Tm1}
\end{equation}

It is not easy calculate the integral in the previous expression,
because it should be compute in the Minkowski space. The solution
of this problem is obtained through a 90$^{\text{o}}$ rotation in
the complex plane $k_{0}$, with this transformation the Minkowski
integration is replaced by an Euclidean integration. Such a
rotation is called Wick rotation and is equivalent to a change of
variables
\begin{equation}
k_{0}=iK_{0,}\text{ \ \ \ \ \ \ \ with \ \ \ } K_{0}\text{\ real}
\end{equation}
then
\begin{equation}
d^{4}k=id^{4}K,\text{ \ \ \ \ }k^{2}=-K^{2},\text{ \ with \ }
K^{2}=K_{0}^{2}+ \overrightarrow{K}^{2}.
\end{equation}
and (\ref{Tm1}) takes the form
\begin{equation}
T_{m}=\int \frac{d^{4}K}{\left(2\pi \right) ^{4}}\ln \left(
1+\frac{ m^{2}_{G}}{ K^{2}}\right).
\end{equation}

Differentiating the previous expression in $m^{2}_{G}$ it is
possible to eliminate the logarithm in the integral
\begin{equation}
\frac{dT_{m}}{dm^{2}_{G}}=\int \frac{d^{4}K}{\left(2\pi \right)
^{4} }\frac{ 1}{ K^{2}+m^{2}_{G}}=\int\limits_{0}^{\infty
}\frac{K^{3}dK}{K^{2}+m^{2}_{G}} \int \frac{ d\Omega }{\left(2\pi
\right) ^{4}}, \label{Tm2}
\end{equation}
and considering for the angular integration the expression
\cite{Muta}:
\begin{equation}
\int \frac{d\Omega }{\left(2\pi \right) ^{4}}=\frac{2\pi
^{2}}{\left(2\pi \right) ^{4}},
\end{equation}
it is possible to rewrite (\ref{Tm2}) in the form
\begin{equation}
\frac{dT_{m}}{dm^{2}_{G}}=\frac{2\pi ^{2}}{\left(2\pi \right)
^{4}} \int\limits_{0}^{\infty }\frac{K^{3}dK}{K^{2}+m^{2}_{G}}.
\label{Tm3}
\end{equation}

The momentum integral in (\ref{Tm3}) is divergent, so it is needed
to use dimensional regularization to calculate it. The area of a
$D$-dimensions unitary sphere is

\begin{equation}
S_{D-1}=\frac{2\pi ^{\frac{D}{2}}}{\Gamma (\frac{D}{2})},
\end{equation}
then the integral (\ref{Tm3}) in $D$-dimensions has the form
\cite{Daemi}
\begin{equation}
\frac{dT_{m}}{dm^{2}_{G}}=\frac{S_{D-1}}{\left(2\pi \right) ^{D}}
\int\limits_{0}^{\infty }\frac{K^{D-1}dK}{K^{2}+m^{2}_{G}}=\frac{
m^{2-\varepsilon }_{G}}{\left(4\pi \right) ^{2-\varepsilon
}}\Gamma \left(-1+ \frac{\varepsilon }{2}\right) \label{Tm4}
\end{equation}
where $\varepsilon =4-D$ and in general
\begin{eqnarray}
\Gamma \left(-n+\frac{\varepsilon }{2}\right) &=&\frac{\left(
-1\right) ^{n} }{n!}\left[\frac{1}{\varepsilon }+\Psi \left(
n+1\right) +O\left(\varepsilon \right) \right], \notag \\ \Psi
\left(n+1\right) &=&1+\frac{1}{2}+...+\frac{1}{n}-\gamma, \notag
\\ \Psi \left(1\right) &=&-\gamma \text{ \ \ with \ }\gamma =0.5772
\end{eqnarray}
All the functions in (\ref{Tm4}) are expanded for
$\varepsilon\rightarrow0$, and cutting the pole in $\varepsilon$
the expression (\ref {Tm4}) takes the form,

\begin{equation}
\frac{dT_{m}}{dm^{2}_{G}}=\left(\frac{m_{G}}{4\pi }\right)
^{2}\left(\ln \left(\frac{m_{G}}{4\pi }\right) ^{2}-const \right),
\end{equation}
and the result for the integral in $m^{2}_{G}$ is
\begin{equation}
T_{m}=\frac{1}{64\pi ^{2}}m^{4}_{G}\ln \frac{m^{4}_{G}}{const}.
\end{equation}

Finally, the expression (\ref{V1}) can be written as

\begin{equation}
V\left(\langle G^{2}\rangle \right) =\frac{\langle
G^{2}\rangle}{4} + 12\frac{1}{64\pi ^{2}}m_{G}^{4}\ln
\frac{m_{G}^{4}}{const },
\end{equation}
that considering (\ref{m2}) and (\ref{g2m}) takes the form

\begin{equation}
V\left(\langle G^{2}\rangle \right) =\frac{\langle
G^{2}\rangle}{4} + \frac{3}{128}g^{2}\langle G^{2}\rangle \ln
\frac{g^{2}\langle G^{2}\rangle }{const }.
\end{equation}
which has a minimum for $G^{2}$ different from zero. This implies
that the gluon condensate is generated spontaneously from the zero
condensate vacuum, as for Savvidy models.

\chapter{Generalization of the Theory for arbitrary values of the
$\alpha$ parameter}

The generalization of the proposed theory, for arbitrary values of
the gauge parameter $\alpha$, is explored. In Sec. 4.1, the ansatz
for the Fock space state that generates the desired perturbative
expansion is defined. The proof that the state satisfies the
physical state conditions imposed for the BRST formalism is also
analyzed in this section. In Sec. 4.2 the modifications to the
usual perturbation theory are calculated, and the introduced
parameters fixed in such a way of making explicit the Lorentz
invariance in the propagator and reproducing a previous
proposition \cite{1995}.

\section{Modified Initial State}

In the present section it is considered the physical vacuum state
able to modify the usual perturbative theory in the way proposed
previously \cite{1995}, for arbitrary values of the $\alpha$
parameter. Some needed results of the Kugo and Ojima quantization,
for arbitrary values of the gauge parameter, are also pointed out.

The equations of motion for arbitrary values of $\alpha$, deduced
from the Kugo and Ojima Lagrangian \cite{Kugo,OjimaTex}, are

\begin{eqnarray}
\Box A_\mu ^a\left(x\right) -\left(1-\alpha \right) \partial _\mu
B^a\left(x\right) &=&0, \nonumber \\ \partial ^\mu A_\mu ^a\left(
x\right) +\alpha B^a\left(x\right)&=&0, \nonumber \\ \Box
B^a\left(x\right) =\Box c^a\left(x\right) =\Box \overline{c}
^a\left(x\right) &=&0. \label{mov}
\end{eqnarray}
Only the solution obtained for the gluonic field differs from the
one obtained for $\alpha =1$, considered earlier \cite{PRD,tesis},
and takes the form \cite{Kugo,OjimaTex}{\small
{\setlength\arraycolsep{0.5pt}\begin{eqnarray} A_\mu ^a\left(
x\right) &=&\sum\limits_{\vec{k}}\left(\sum\limits_{\sigma
=1,2}A_{\vec{k},\sigma }^af_{k,\mu }^\sigma \left(x\right)
+A_{\vec{k} }^{L,a}f_{k,L,\mu }\left(x\right) +B_{\vec{k}}^a\left[
f_{k,S,\mu }\left(x\right) +(1-\alpha )\partial _\mu
D_{(x)}^{(1/2)}g_k(x)\right]\right)\nonumber\\ &&\qquad \ \ \
+h.c., \label{sol1}
\end{eqnarray}}}\noindent where the wave packets in (\ref{sol1})
are taken as
\begin{eqnarray}
g_k\left(x\right) &=&\frac 1{\sqrt{2Vk_0}}\exp \left(
-ikx\right),\quad kx=k_0 x_0- \vec{k}\vec{x}, \ \ k_0=\left|
\vec{k} \right| \nonumber \\ f_{k,\mu }^\sigma \left(x\right)
&=&\frac 1{\sqrt{2Vk_0}}\epsilon _\mu ^\sigma \left(k\right) \exp
\left(-ikx\right). \label{paq}
\end{eqnarray}

The polarization vectors $\epsilon _\mu ^\sigma \left(k\right)$,
$\epsilon _{L,\mu }\left(k\right)$, $\epsilon _{S,\mu }\left(
k\right)$ are defined by
\begin{eqnarray}
\vec{k}\cdot \vec{\epsilon^\sigma} \left(k\right) &=&0,\ \epsilon
_0^\sigma \left(k\right) =0,\ \vec{\epsilon^\sigma} \left(
k\right) \cdot \vec{ \epsilon^\tau} \left(k\right) =\delta
^{\sigma \tau },\text{ for }\sigma,\tau =1,2 \nonumber \\ \epsilon
_{L,\mu }\left(k\right) &=&-ik_\mu =-i\left(\left|
\vec{k}\right|,-\vec{k}\right),\  \nonumber \\ \epsilon _{S,\mu
}\left(k\right) &=&-i\frac{\overline{k}_\mu }{2\left| \vec{
k}\right| ^2}=\frac{-i\left(\left| \vec{k}\right|,\vec{k}\right)
}{2\left| \vec{k}\right| ^2}. \label{polar}
\end{eqnarray}

The commutation relations between the creation and annihilation
operators are given by

\begin{equation}
\begin{array}{cccccc}
& A_{\vec{k}^{\prime },\sigma ^{\prime }}^{a^{\prime }+} &
A_{\vec{k} ^{\prime }}^{L,a^{\prime }+} & B_{\vec{k}^{\prime
}}^{a^{\prime }+} & c_{ \vec{k}^{\prime }}^{a+} &
\overline{c}_{\vec{k}^{\prime }}^{a+} \\ A_{\vec{k},\sigma }^a &
\delta ^{aa^{\prime }}\delta _{\vec{k}\vec{k} ^{\prime }}\delta
_{\sigma \sigma ^{\prime }} & 0 & 0 & 0 & 0 \\ A_{\vec{k}}^{L,a} &
0 & 0 & -\delta ^{aa^{\prime }}\delta _{\vec{k}\vec{k} ^{\prime }}
& 0 & 0 \\ B_{\vec{k}}^a & 0 & -\delta ^{aa^{\prime }}\delta
_{\vec{k}\vec{k}^{\prime }} & 0 & 0 & 0 \\ c_{\vec{k}}^a & 0 & 0 &
0 & 0 & i\delta ^{aa^{\prime }}\delta _{\vec{k}\vec{k }^{\prime }}
\\ \overline{c}_{\vec{k}}^a & 0 & 0 & 0 & -i\delta ^{aa^{\prime
}}\delta _{\vec{ k}\vec{k}^{\prime }} & 0
\end{array}
\label{commu}
\end{equation}

The integro-differential operator $D_{(x)}^{(1/2)}$, is defined by

\begin{equation}
D_{(x)}^{(1/2)}\equiv 1/2(\nabla
^2)^{-1}(x_0\partial _0-1/2),
\end{equation}
and works as an ``inverse'' of the d'Alembertian for simple pole
functions \cite{Nakanishi}

\begin{equation}
\Box D_{(x)}^{(1/2)}f\left(x\right) =f\left(x\right) \text{ \quad
if \ \ } \Box f\left(x\right) =0.
\end{equation}

The value of $\partial _\mu D_{(x)}^{(1/2)}g_k(x)$ in (\ref{sol1})
is determined with the use of the wave packets (\ref{paq}) and
polarization vectors (\ref{polar}) considered. The result obtained
is

\begin{eqnarray}
\partial _\mu D_{(x)}^{(1/2)}g_k(x) &=&\partial _\mu 1/2(\nabla
^2)^{-1}(x_0\partial _0-1/2)\frac 1{\sqrt{2Vk_0}}\exp \left(
-ikx\right) \nonumber \\ &=&A\left(k_0,x_0\right) f_{k,L,\mu
}\left(x\right) +B_\mu \left(k,x\right), \label{1}
\end{eqnarray}
where $A\left(k_0,x_0\right)$ and $B_\mu \left(k,x\right)$ are
defined by:

\begin{eqnarray*}
A\left(k_0,x_0\right) &\equiv &\frac 1{2\left| \vec{k} \right|
^2}\left(ik_0x_0+1/2\right), \\ B_\mu \left(k,x\right) &\equiv
&\frac{ik_0\delta _{\mu 0}}{2\left| \vec{k} \right| ^2}g_k(x).
\end{eqnarray*}

In similar way for $\partial _{\mu }D_{(x)}^{(1/2)}g_{k}^{\ast
}(x)$ is obtained:

\begin{equation}
\partial _\mu D_{(x)}^{(1/2)}g_k^{*}(x)=A^{*}\left(k_0,x_0\right)
f_{k,L,\mu }^{*}\left(x\right) +B_\mu ^{*}\left(k,x\right).
\label{2}
\end{equation}

Now it is possible to define the proposition for the QCD vacuum
state. The considered one is a generalization of the previously
proposed \cite{PRD,tesis}, in which the constant accompanying the
non-physical sector has allowed to be gauge dependent in order to
obtain a covariant gluon propagator. This state is physically
equivalent to the one proposed in the previous work
\cite{PRD,tesis}, because its physical part (transverse gluons
part) was not modified. The modifications were introduced only in
the non-physical sector, as usual for this sector, to make
explicit the covariance.

The new vacuum state is defined by the expression

\begin{equation}
\mid\Psi \rangle =\exp \left\{
\sum\limits_{a=1}^8\sum\limits_{\vec{p} _i,\left| \vec{p}_i\right|
<P}\left[ \sum\limits_{\sigma =1,2}\frac{C_\sigma \left(P\right)
}2A_{\vec{p}_i,\sigma }^{a+}A_{\vec{p}_i,\sigma }^{a+}+C_3\left(
\alpha,P\right) \left(B_{\vec{p}_i}^{a+}A_{\vec{p}
_i}^{L,a+}+i\overline{c}_{\vec{p}_i}^{a+}c_{\vec{p}_i}^{a+}\right)
\right] \right\} \mid 0\rangle. \label{Vacuum1}
\end{equation}
In this state the gluon pairs are created with very small momenta
$\vec{p}_i$, but different from zero, with the objective of
avoiding the difficulties appearing if they are considered with
zero momentum directly. The sum over $\vec{p}_i$ for $\left|
\vec{p}_i\right| <P$\ is introduced to avoid any preferential
direction in the space when the limit $P \rightarrow 0$ is
considered, taking into account that the volume $V \rightarrow
\infty$. Is interesting to note that this sum was not introduced
in a previous work \cite{PRD,tesis}, for the $\alpha =1$ analysis,
because the operator $D_{(x)}^{(1/2)}$ was not involved and the
results were independent from the introduced auxiliary momentum
direction.

In the analysis following the work will be explicitly done with a
generic color and a generic momentum $\vec{p}_i$, for simplifying
the exposition. This can be done because in the free theory all
different colors and momenta can be worked out independently,
thanks to the commutation relations defined (\ref{commu}). At the
necessary point of the analysis all the contributions will be
included.

The proof that the proposed state (\ref{Vacuum1}) satisfies the
required physical state conditions,

\begin{eqnarray}
&&Q_B\mid \Psi \rangle =0, \nonumber \\ &&Q_C\mid \Psi \rangle =0,
\label{QBQC}
\end{eqnarray}
is exactly the same advanced in the previous work, because the
constant $C_3\left(\alpha,P\right)$ does not make any change. So
this state is a physical state of the theory and it is possible to
state that the non-physical sector of $\mid \Psi \rangle$ is
undetectable in the physical world and it can be neglect in
physical calculations, like in the calculation of the norm for
example.

Then for the calculation of the norm a similar result, to the one
in \cite{PRD,tesis}, is obtained

\[
N=\langle \Psi \mid \Psi \rangle =\prod\limits_{\sigma
=1,2}\prod\limits_{a=1}^8\prod\limits_{\vec{p}_i, \left|
\vec{p}_i\right| <P} \left[ \sum\limits_{m=0}^\infty \left|
C_\sigma \left(P\right) \right| ^{2m}\frac{\left(2m\right)
!}{\left(m!\right) ^2}\right], \text{\quad for }\left| C_\sigma
\left(P\right) \right| <1
\]

The normalized physical vacuum state is then defined by

\begin{equation}
\mid \tilde{\Psi}\rangle =\frac 1{\sqrt{N}}\mid \Psi \rangle.
\end{equation}

\section{Modified Perturbation Theory}

In the present section the modification to the usual perturbation
theory is calculated. It is introduced through the modification of
the usual vacuum state (empty of the Fock space), by the state
proposed in the previous section (\ref{Vacuum1}). The analysis is
generalized for arbitrary values of the gauge parameter $\alpha$.
At the end a brief analysis of the $\alpha$ generalization
consequences, on the previous calculations for $\langle G^2
\rangle$, quark and gluon masses, is done.

\newpage

As it was pointed out in a previous work \cite{PRD,tesis}, the
modification to the gluon generating functional of the
free-particle Green functions (and to the propagator also) is
determined by:
\begin{equation}
\langle \widetilde{\Psi }\mid \exp \left\{ i\int
d^4x\sum_{a=1,..8}J^{\mu,a}\left(x\right) A_\mu ^{a-}\left(
x\right) \right\} \exp \left\{ i\int d^4x\sum_{a=1,..8}
J^{\mu,a}\left(x\right) A_\mu ^{a+}\left(x\right) \right\} \mid
\widetilde{\Psi }\rangle. \label{mod}
\end{equation}

Considering the expressions (\ref{sol1}), (\ref{1}) and (\ref{2}),
the gluon annihilation and creation operators in (\ref{mod}) have
the form {\small
\begin{eqnarray}
A_\mu ^{a+}\left(x\right) &=&\sum\limits_{\vec{k}}\left(
\sum\limits_{\sigma =1,2}A_{\vec{k},\sigma }^af_{k,\mu }^\sigma
\left(x\right) +A_{\vec{k}}^{L,a}f_{k,L,\mu }\left(x\right)
+\right. \nonumber \\ &&\text{ \qquad }\left. +B_{\vec{k}}^a\left[
f_{k,S,\mu }\left(x\right) +(1-\alpha)\left(A\left(k_0,x_0\right)
f_{k,L,\mu }\left(x\right) +B_\mu \left(k,x\right) \right) \right]
\right), \nonumber \\ A_\mu ^{a-}\left(x\right)
&=&\sum\limits_{\vec{k}}\left(\sum\limits_{\sigma
=1,2}A_{\vec{k},\sigma }^{a+}f_{k,\mu }^{\sigma *}\left(x\right)
+A_{\vec{k}}^{L,a+}f_{k,L,\mu }^{*}\left(x\right) +\right..
\nonumber \\ &&\text{ \qquad }\left. +B_{\vec{k}}^{a+}\left[
f_{k,S,\mu }^{*}\left(x\right)
+(1-\alpha)\left(A^{*}\left(k_0,x_0\right) f_{k,L,\mu }^{*}\left(
x\right) +B_\mu ^{*}\left(k,x\right) \right) \right] \right).
\label{4.11}
\end{eqnarray}}

As it can be noticed in the expressions (\ref{Vacuum1}) and
(\ref{4.11}) the transverse and ghost terms are not modified with
the generalization of the theory for arbitrary values of $\alpha$.
Then their contributions are the same calculated previously
\cite{PRD,tesis}. It is important to remark that the calculations
are done decomposing the expression (\ref{mod}) in products of
each mode contribution, thanks to commutation relations
(\ref{commu}).

The result obtained for the transverse mode contributions, for
$\left| C_\sigma \left(P\right) \right|<1$, is \cite{PRD,tesis}
{\small
\begin{eqnarray}
&&\langle 0\mid \exp \left(\sum_{\sigma =1,2}\frac{C_\sigma
^{*}\left(P\right) }2A_{\vec{p}_i,\sigma }^aA_{\vec{p}_i,\sigma
}^a\right) \exp \left\{ i\int d^4xJ^{\mu,a}\left(x\right)
\sum\limits_{\sigma =1,2}A_{\vec{ p}_i,\sigma }^{a+}f_{p_i,\mu
}^{\sigma *}\left(x\right) \right\} \nonumber
\\ &&\times \exp \left\{ i\int d^4xJ^{\mu,a}\left(x\right)
\sum\limits_{\sigma =1,2}A_{\vec{p}_i,\sigma }^af_{p_i,\mu
}^\sigma \left(x\right) \right\} \exp \left(\sum_{\sigma
=1,2}\frac{C_\sigma \left(P\right) }2A_{\vec{p}_i,\sigma
}^{a+}A_{\vec{p}_i,\sigma }^{a+}\right) \mid 0\rangle \nonumber
\\ &&=\exp \left\{ -\sum\limits_{\sigma =1,2}\left(J_{p_i,\sigma
}^a\right) ^2 \frac{\left(C_\sigma \left(P\right) +C_\sigma
^{*}\left(P\right) +2\left| C_\sigma \left(P\right) \right|
^2\right) }{2\left(1-\left| C_\sigma \left(P\right) \right|
^2\right) }\right\}, \label{Transv}
\end{eqnarray}}\noindent where the normalization factor has been
canceled, and it should be reminded that the momenta $\vec{p}_i$
are very small and the sources are located in a space finite
region. The following simplified notation was also introduced
\cite{PRD,tesis}
\[
J_{p_i,\sigma }^a=\int \frac{d^4x}{\sqrt{2Vp_{i0}}}J^{\mu,a}\left(
x\right) \epsilon _{\sigma,\mu }\left(p_i\right).
\]

The calculation of longitudinal and scalar mode contributions is
in this occasion more elaborated, and a brief exposition of it
could be found in the Appendix B, the result obtained for $\left|
C_3\left(\alpha,P\right) \right|<1$ is

 {\small
\begin{eqnarray}
\exp \left\{ -\int \frac{d^4xd^4y}{2Vp_{i0}}J^{\mu,a}\left(
x\right) J^{\nu,a}\left(y\right) \left[ \left(\frac{C_3\left(
\alpha,P\right) +C_3^{*}\left(\alpha,P\right) +2\left| C_3\left(
\alpha,P\right) \right| ^2}{\left(1-\left| C_3\left(\alpha
,P\right) \right| ^2\right) }\right)\times \right. \right.&&
\nonumber \\ \times \left. \left. \left(\epsilon _{S,\mu }\left(
p_i\right) \epsilon _{L,\nu }\left(p_i\right) +\frac{\left(
1-\alpha \right) }{4\left| \vec{p}_i\right| ^2} \left[ \epsilon
_{L,\mu }\left(p_i\right) +i2p_{i0}\delta _{\mu 0}\right] \epsilon
_{L,\nu }\left(p_i\right) \right) \right] \right\},&&
\label{LonSca}
\end{eqnarray}}\noindent the normalization factor has been
canceled.

Now substituting the expressions (\ref{Transv}) and (\ref{LonSca})
in (\ref{mod}). Performing some algebraic manipulations, keeping
in mind the properties of the polarization vectors defined
(\ref{polar}), considering $C_1\left(P\right) =C_2\left(
P\right)$, and finally introducing the contributions for all
momenta $\vec{p}_i$ for $\left| \vec{p}_i\right| <P$, the
following result is obtained

{\small
\begin{eqnarray}
&&\exp \left\{ \int \frac{d^4xd^4y}{2V}J^{\mu,a}\left(x\right)
J^{\nu,a}\left(y\right) \sum\limits_{\vec{p}_i, \left|
\vec{p}_i\right| <P} \frac 1{p_{i0}}\left[ \left(\frac{C_1\left(
P\right) +C_1^{*}\left(P\right) +2\left| C_1\left(P\right) \right|
^2}{2\left(1-\left| C_1\left(P\right) \right| ^2\right) }\right)
g_{\mu \nu }+\right. \right. \nonumber
\\ && \ \ +\left(\frac{C_3\left(\alpha,P\right)
+C_3^{*}\left(\alpha,P\right) +2\left| C_3\left(\alpha,P\right)
\right| ^2}{\left(1-\left| C_3\left(\alpha,P\right) \right|
^2\right) }-\frac{C_1\left(P\right) +C_1^{*}\left(P\right)
+2\left| C_1\left(P\right) \right| ^2}{ \left(1-\left| C_1\left(
P\right) \right| ^2\right) }\right) \frac{\bar{p} _{i\mu }p_{i\nu
}}{2\left| \vec{p}_i\right| ^2} \nonumber \\ && \ \ \left. \left.
+ \left(\frac{C_3\left(\alpha,P\right) +C_3^{*}\left(\alpha
,P\right) +2\left| C_3\left(\alpha,P\right) \right| ^2}{\left(
1-\left| C_3\left(\alpha,P\right) \right| ^2\right) }\right)
\frac{\left(1-\alpha \right) }{4\left| \vec{p}_i\right| ^2}\left(
p_{i\mu }p_{i\nu }-2p_{i0}\delta _{\mu 0}p_{i\nu }\right) \right)
\right\}. \label{mod1}
\end{eqnarray}}\noindent It is interesting to note at this point
that the combinations of $C_1\left(P\right)$ and $C_3\left(\alpha
,P\right)$, in the expression (\ref{mod1}), are real. So even if
these parameters are complex the modifications introduced by them
are real. Then, after knowing that complex parameters essentially
produce the same result as the real ones, they will be considered
as real in what follows.

In the expression (\ref{mod1}) the addition is changed by an
integration, considering that $V \rightarrow \infty$, in
accordance with the relation

\begin{equation}
\frac 1V \sum\limits_{\vec{p}_i, \left| \vec{p}_i\right| <P}=\frac
1{\left(2\pi \right) ^3}\int\limits_0^P dp\ p^2\int\limits_0^\pi
\sin\theta d\theta \int\limits_0^{2\pi }d\varphi, \label{integr}
\end{equation}
and the integrals are calculated.

After that, the parameters introduced were considered behaving in
following form, for $P\sim 0$,
\begin{eqnarray}
C_1\left(P\right) &\sim &1-\frac{C_1}2P^2,\ C_1>0, \nonumber \\
C_3\left(\alpha,P\right) &\sim &1-\frac{C_3\left(\alpha \right)
}2P^2,\ C_3\left(\alpha \right) >0.  \label{cond}
\end{eqnarray}

The result obtained for the expression (\ref{mod1}), in the limit
$P\rightarrow 0$, is then

\begin{eqnarray}
&&\exp \left\{ \int d^4xd^4yJ^{\mu,a}\left(x\right) J^{\nu
,a}\left(y\right) \left[ \frac{g_{\mu \nu }}{\left(2\pi \right)
^2C_1}+\frac 1{3\left(2\pi \right) ^2}\left(\frac 1{C_3\left(
\alpha \right) }-\frac 1{C_1}\right) \left(g_{\mu \nu }+2\delta
_{\mu 0}\delta _{\nu 0}\right) \right. \right. \nonumber \\
&&\qquad\qquad\qquad\qquad\qquad\qquad\qquad \ \ \left. \left.
-\frac{\left(1-\alpha \right) }{\left(2\pi \right) ^26C_3\left(
\alpha \right) }\left(g_{\mu \nu }+2\delta _{\mu 0}\delta _{\nu
0}\right) \right] \right\}. \label{mod2}
\end{eqnarray}

The constant $C_3\left(\alpha \right)$ is fixed in order to cancel
the terms proportional to $\delta _{\mu 0}\delta _{\nu 0}$, to
make explicit the Lorentz covariance. That's why this non-physical
part was introduced in the vacuum state. Then $C_3\left(\alpha
\right)$ is determined by the expression
\[
C_3\left(\alpha \right) =\frac{\left(1+\alpha \right)}{2} C_1,
\]
and (\ref{mod2}) takes the form
\begin{equation}
\exp \left\{ \int d^4xd^4yJ^{\mu,a}\left(x\right) J^{\nu,a}\left(
y\right) \frac{g_{\mu \nu }}{\left(2\pi \right) ^2} \frac 1{C_1}
\right\}. \label{mod3}
\end{equation}
In the expression (\ref{mod3}) is important to note that the term
in the exponential is real and non-negative (\ref{cond}). And that
it is independent of the gauge parameter, which is a very
interesting result and it was assumed in a previous work
\cite{1995}.

Considering the above remarks and defining in the generating
functional modification an alternative nonnegative constant
$C=\frac{2\left(2\pi \right)^2}{C_1}$ (which will be called
condensate parameter) for further convenience, the expression
(\ref{mod3}) takes the form

\[
\exp \left\{ \int d^4xd^4y\sum\limits_{a=1}^8J^{\mu,a}\left(
x\right) J^{\nu,a}\left(y\right) \frac{g_{\mu \nu }C}{\left(2\pi
\right) ^42} \right\}.
\]

For the above construction the vacuum in the zero momentum limit
has the form {\small
\begin{equation}
\mid \Psi \rangle =\exp \left[ \sum\limits_{a=1}^8\frac
12A_{0,1}^{a+}A_{0,1}^{a+}+\frac
12A_{0,2}^{a+}A_{0,2}^{a+}+B_0^{a+}A_0^{L,a+}+i\overline{c}
_0^{a+}c_0^{a+}\right] \mid 0\rangle.\label{Vacuum3}
\end{equation}}

\noindent Then in the zero momentum limit this state has exactly
the same form of the one proposed previously \cite{PRD,tesis}, the
only difference appearing is introduced in the way in which the
zero momentum limit is taken. That previous work \cite{PRD,tesis}
corresponds to the particular case in which $\alpha=1$, $C_3\left(
\alpha=1 \right) =C_1$.

The modification to the ghost generating functional of the
free-particle Green functions is the same calculated previously
\cite{PRD,tesis}. It is because the ghost sector in the vacuum
state remains unchanged. As was mentioned there, for the value
$C_3\left(\alpha,0\right)=1$ as was fixed here, there is no
modification of the ghost sector and its generating functional and
propagator remain the same of the usual PQCD. The consequence of
this unchanged sector was mentioned in Section 3.2, that is it
guarantees the fulfillment of a Ward identity in the one loop
calculation for the gluon self energy.

Then, after all the previous analysis, it is concluded that the
vacuum state introduced (\ref{Vacuum3}) modifies the usual
perturbation theory only through a change in the gluon generating
functional of the free-particle Green functions in such a way that
the total gluon propagator (including the usual perturbative
piece) takes the form

\begin{equation}
\widetilde{D}_{\mu \nu }^{ab}(x-y)=\int \frac{d^4k}{\left(2\pi
\right) ^4} \delta ^{ab}\left[ \frac 1{k^2}\left(g_{\mu \nu
}-\left(1-\alpha \right) \frac{k_\mu k_\nu }{k^2}\right) -iC\delta
\left(k\right) g_{\mu \nu }\right] \exp \left\{ -ik\left(
x-y\right) \right\}.  \label{propag}
\end{equation}
in which the first term is the usual perturbative propagator.

Finally it is possible to analyze the consequences of the present
generalization on previous calculations, for $\langle G^2
\rangle$, quark and gluon masses, done in Chapter 3.

For the mean value of $G^2$ and the quarks masses, after
performing calculations like the ones realized previously, was
found that the results are exactly the same obtained before,
independently of the $\alpha$ parameter values. However for the
gluon mass is obtained the following result

\[
m_G^2=-\frac{3\left(1+\alpha \right) g^2C}{\left(2\pi \right) ^4}.
\]

This result, $\alpha $ dependent, does not represent a
contradiction because the gluon mass is tachyonic for all the
admissible values of the $\alpha$ parameter (the non-negative
ones), then it is not a physical magnitude and could depend on the
gauge choice. However at the present state of the analysis the
general proof of the gauge invariance of the theory is lacking and
will be considered in future works.

\chapter{Summary}

This work considers the extension of the former ones
\cite{1995}-\cite{tesis}. The results obtained can be summarized
as follows:

1) The elements of the modified perturbative expansion for QCD
initiated in \cite{1995} were specified. The diagram technique
embodies non-perturbative information associated to a non-trivial
vacuum in which exists a condensate of soft gluon pairs, as found
in \cite{PRD,tesis}.

2) The former calculation of the mean value of $G^{2}$ was
completed by including all corrections of order $g^{2}$,
disregarding the terms not including the condensate effects. The
result was the same obtained previously in the tree approximation
\cite{1995,PRD}. The mean value of $G^{2}$ is a relevant quantity
for the description of the QCD vacuum \cite{Zakharov}, and has
been calculated previously considering non-perturbative methods.
The condensate parameter $g^2C$ was determined to reproduce the
accepted value of $\langle g^2 G^2 \rangle$.

3) The contribution of the condensate to the gluon self-energy, up
to the $g^2$ order, was calculated. A tachyonic result for the
gluon mass was obtained, for the parameter values employed in the
construction of the initial state defined in \cite{PRD,tesis}.
This because the condensate parameter $C$ was determined as real
and non-negative in that work. The polarization of the tachyonic
mode is transverse. As remarked before there are recent arguments
aiming at the possibility of the existence of a tachyonic gluon
mass \cite{tachyon1,tachyon2}, and its role in improving the quark
antiquark potential in bound state models of mesons
\cite{tachyon2}.

4) An evaluation for the quark masses, under the influence of the
condensate, was done. Various solutions were obtained for the
masses of these particles, among which only one is real for all
the values the current mass parameter. Employing this particular
solution some interesting predictions were derived. In particular
values of near one third of the nucleon mass, for the light quark
masses, were obtained. Further, the effective quark masses were
evaluated as determined by the known current mass for each flavor.
By mean of these values, the ground state energies of some hadron
resonances, reported in \cite{Report}, were estimated through the
simple addition of the masses of their known constituent quarks.
It could be remarked that the results obtained are reasonable good
ones in comparison with the experimental data, after taking into
account the high errors in the experimentally determined
Lagrangian masses. This makes possible to sustain the hypothesis
about the almost glue nature of the constituent mass values of the
light quarks. The small contribution of the binding potential
energy of such resonances is also indicated. This conclusion then
supports the applicability of non-relativistic approximations in
their study.

5) The effective potential was determined, in the approximation
considered in a previous work \cite{1995}, as a function of
$\langle G^2 \rangle$. As it was obtained previously \cite{1995}
the condensate is generated spontaneously from the zero field
vacuum, as in the Savvidy Model. That is, it was shown that the
effective action has a minimum for a mean value of $G^2$ different
from zero.

It is worth resuming also the main aspects of the modified
expansion that have been clarified after the initial work
\cite{1995}:

A) The Feynman expansion, depending on the condensate parameter
and a gauge parameter $\alpha =1$, corresponds to the Wick
expansion around a physical state of the free QCD.

B) Also in \cite{PRD,tesis} the condensation parameter $C$ was
defined as real and positive. This result determined the tachyonic
value for the gluon mass evaluated here and the masses for up and
down quarks as 1/3 of the nucleon mass.

C) Another aspect which is important to underline is that the
selection of the parameters in the initial state \cite{PRD,tesis}
was designed also to impose the absence of a modification for the
free ghost propagator. However, as it followed from the present
analysis, this property in turns is related with the fact that a
condensate modification of the ghost propagator can produce a
longitudinal contribution to the self-energy. But, such a term
should not exist for the transversality condition of the
polarization operator (Ward identity) to be obeyed. Thus, its
appearance could break manifestly the gauge invariance. Therefore,
the present work is also given foundation to the non-modified
ghost propagator choice considered in \cite{1995}-\cite{tesis}.

5) Finally, it was explored the generalization of the theory for
arbitrary values of the $\alpha$ gauge parameter. A particular
vacuum state for a Perturbative QCD was proposed, it is formed by
a coherent superposition of zero momentum gluon pairs, and it is a
generalization of the one proposed in a previous work
\cite{PRD,tesis}. It was analyzed that the proposed state
satisfies the BRST physical conditions imposed by the operational
quantization method for gauge fields, developed by Kugo and Ojima
\cite{Kugo,OjimaTex}. The modifications in the perturbation theory
introduced by the proposed vacuum state were calculated for
arbitrary values of the $\alpha$ gauge parameter. The parameters
introduced in the definition of the vacuum state were fixed in
such a way that they reproduce the modification to the gluon
propagator initially proposed and based in a previous work
\cite{1995} starting from a functional analysis. The condensation
parameter $C$, that modifies the standard propagator multiplying a
term containing a delta function in the four-momentum $p$, was
determined to be real and nonnegative. The consequences of the
generalized theory on previously calculated magnitudes were
analyzed. For the calculation of the mean value of $G^2$ and the
quark masses was obtained that the results are identical to the
ones determined for $\alpha=1$. The expression for the gluon mass
was also calculated and it is determined to be gauge dependent and
tachyonic for all the values of the $\alpha$ parameter. The
general proof of the gauge invariance of the theory is lacking and
will be study later on.

At the present stage some possible lines for the extension of the
work can be visualized:

1) To explore in more detail the various possibilities for the
quantization of the zero modes of the gluon field. It could allow
to bypass the employment of the zero momentum limit used in the
construction of the initial state, and also perhaps permit to
simplify its introduction and the study of alternatives for the
modified expansion describing different physical ground states.

2) Improve the study of the effective action as a functional of
the condensate parameter, in order to search for a variant of the
leading logarithm model useful for the investigation of field
configurations associated to inter-quark strings, nucleon-nucleon
potentials, etc.

3) Start the introduction of the field and coupling constant
renormalization in the scheme. This would allow to investigate if
the infrared behavior of the running coupling of the PQCD can be
regularized by the modified expansion. The evolution of the masses
at high energy will be another task to be worth considering. At
last the determination of the absolute value of the constant $C$
would be also possible after finding the running coupling
evolution.

4) To investigate the possibility for a derivation of existing
successful bounded state models for the heavy quark mesons
\cite{sommer1}-\cite{El-Hady}, as realized by the a ladder
approximation for the Bethe-Salpeter equation within the proposed
modified expansion. The presence of a tachyonic gluon propagator
in the approach (which is argued to have the effect of introducing
a linearly rising component in the inter-quark potential
\cite{tachyon2}) and the obtained constituent values for the light
quark masses already support such a possibility.

\appendix

\chapter{Diagrams for the evaluation of $\langle
G^{2}\rangle$}

For the exposition of diagrams appearing in the $\left\langle
G^{2}\right\rangle$ calculation, only the gluon and quarks will be
taken into account. The terms related with ghosts do not
contribute. This conclusion can be extracted easily following the
diagrams related with quarks, since the associated ghost diagrams
are similar. All the contributions to $\left\langle G^{2}
\right\rangle$ of diagrams not including the condensate parameter
are disregarded, as was justified in the introduction.

Expanding the exponential in the vertices in (\ref{funct2}), up to
order $g^{2}$ for the gluon and quark contributions it is obtained
{\small {\setlength\arraycolsep{0.1pt}
\begin{eqnarray}
Z\left[ J,\bar{\xi},\xi \right]&=&N\left[
1+\frac{S_{abc}^{G}}{3!i^{2}} \frac{\delta }{\delta
J_{a}}\frac{\delta }{\delta J_{b}}\frac{\delta }{ \delta
J_{c}}+\frac{S_{abcd}^{G}}{4!i^{3}}\frac{\delta }{\delta
J_{a}}\frac{ \delta }{\delta J_{b}}\frac{\delta }{\delta
J_{c}}\frac{\delta }{\delta J_{d}
}+\frac{S_{iaj}^{Q}}{i^{2}}\frac{\delta }{\delta
\bar{\xi}_{i}}\frac{\delta }{\delta J_{a}}\frac{\delta }{\delta
\left(-\xi _{j}\right) }\right. \nonumber \\ &&+
\frac{1}{2}\left(\frac{S_{abc}^{G}}{3!i^{2}}\frac{\delta }{\delta
J_{a}}\frac{\delta }{\delta J_{b}}\frac{\delta }{\delta
J_{c}}\frac{ S_{def}^{G}}{3!i^{2}}\frac{\delta }{\delta
J_{d}}\frac{\delta }{\delta J_{e}} \frac{\delta }{\delta
J_{f}}+\frac{S_{iaj}^{Q}}{i^{2}}\frac{\delta }{\delta
\bar{\xi}_{i}}\frac{\delta }{\delta J_{a}}\frac{\delta }{\delta
\left(-\xi _{j}\right) }\frac{S_{kbl}^{Q}}{i^{2}}\frac{\delta
}{\delta \bar{\xi}_{k}} \frac{\delta }{\delta J_{b}}\frac{\delta
}{\delta \left(-\xi _{l}\right) } \right. \nonumber \\ &&\qquad \
\ \ \left. +\left. 2\frac{S_{abc}^{G}}{3!i^{2}}\frac{\delta
}{\delta J_{a}}\frac{\delta }{\delta J_{b}}\frac{\delta }{\delta
J_{c}}\frac{ S_{idj}^{Q}}{i^{2}}\frac{\delta }{\delta
\bar{\xi}_{i}}\frac{\delta }{\delta J_{d}}\frac{\delta }{\delta
\left(-\xi _{j}\right) }\right) \right] Z_{0} \left[
J,\bar{\xi},\xi \right], \label{des2}
\end{eqnarray}}}
\noindent where the norm ($N$) up to order $g^2$ is given by the
expression {\small {\setlength\arraycolsep{0.1pt}
\begin{eqnarray}
N &=&1-\left\{ \left[ \frac{S_{abcd}^{G}}{4!i^{3}}\frac{\delta
}{\delta J_{a} }\frac{\delta }{\delta J_{b}}\frac{\delta }{\delta
J_{c}}\frac{\delta }{ \delta J_{d}}\right. \right. \nonumber \\
&&\qquad +\frac{1}{2}\left(
\frac{S_{abc}^{G}}{3!i^{2}}\frac{\delta }{\delta
J_{a}}\frac{\delta }{\delta J_{b}}\frac{\delta }{\delta
J_{c}}\frac{ S_{def}^{G}}{3!i^{2}}\frac{\delta }{\delta
J_{d}}\frac{\delta }{\delta J_{e}} \frac{\delta }{\delta
J_{f}}+\frac{S_{iaj}^{Q}}{i^{2}}\frac{\delta }{\delta
\bar{\xi}_{i}}\frac{\delta }{\delta J_{a}}\frac{\delta }{\delta
\left(-\xi _{j}\right) }\frac{S_{kbl}^{Q}}{2!i^{2}}\frac{\delta
}{\delta \bar{\xi}_{k}} \frac{\delta }{\delta J_{b}}\frac{\delta
}{\delta \left(-\xi _{l}\right) } \right. \nonumber \\ &&\qquad
\qquad \left. \left. \left.
+2\frac{S_{abc}^{G}}{i^{2}}\frac{\delta }{\delta
J_{a}}\frac{\delta }{\delta J_{b}}\frac{\delta }{\delta
J_{c}}\frac{ S_{idj}^{Q}}{2!i^{2}}\frac{\delta }{\delta
\bar{\xi}_{i}}\frac{\delta }{ \delta J_{d}}\frac{\delta }{\delta
\left(-\xi _{j}\right) }\right) \right] Z_{0}\left[
J,\bar{\xi},\xi \right] \right\} _{J,\bar{\xi},\xi =0}.
\label{des3}
\end{eqnarray}}}

Now, the diagrams generated by the action of the functional
differential operator in (\ref{des1}) on (\ref{des2}), after
taking into account (\ref{des3}), are analysed.

The third term in the expansion (\ref{des1}) is already of order
$g^{2}$. Up to this order also, it will only contribute with the
first term in the expansion (\ref{des2}), which only generates the
diagram shown in Fig. \ref{G21}.

\begin{figure}[h]
\begin{center}
\includegraphics[scale=0.15,angle=0]{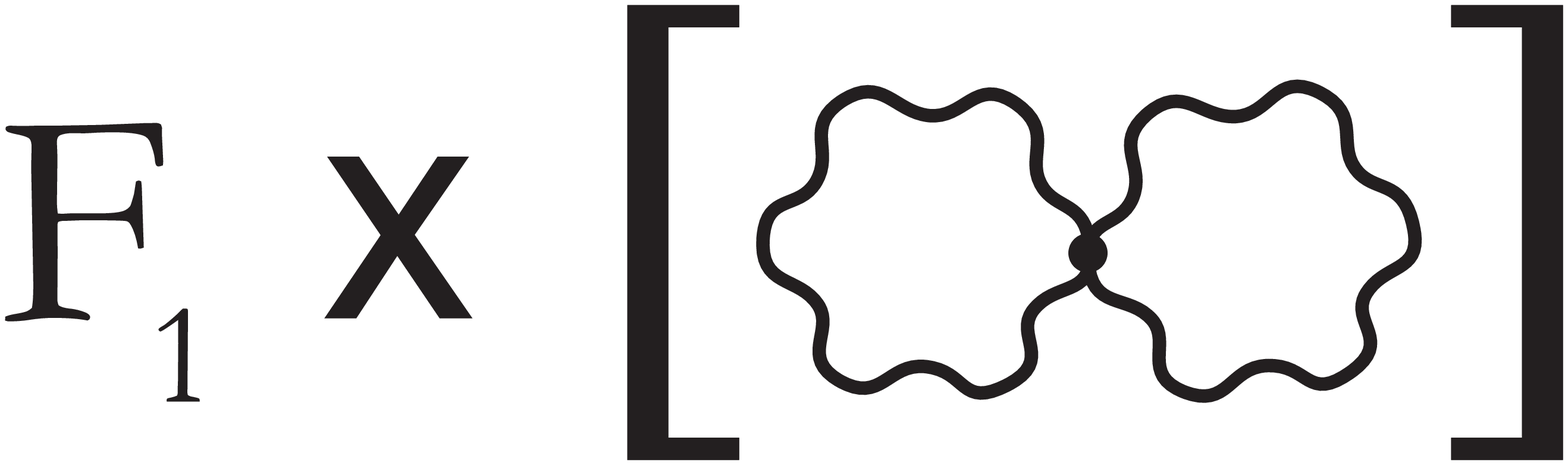}
\end{center}
\vspace{-0.7cm} \caption{Diagram 1.} \label{G21}
\end{figure}

Substituting the expressions for propagators and vertices in the
diagram of the Fig. \ref{G21}, disregarding the terms in which the
condensation parameter $C$ do not appear and considering that the
integrals of the form

\begin{equation}
I=\int \frac{d^{D}q}{\left(-q^{2}\right) ^{\alpha }}=0\qquad
\text{for \ } \alpha >0 \label{regdim}
\end{equation}
within dimensional regularization, the following result for the
first diagram follows:

\begin{equation}
-\frac{1}{8}\left[ \frac{576g^{2}C^{2}}{\left(2\pi \right)
^{8}}\int d^{4}x \right], \label{D1}
\end{equation}
where $\int d^{4}x$ means the space-time volume. The symmetry
factor of (\ref{regdim}) is $F_{1}=-\frac{1}{8}$. This diagram
represents the tree approximation for the evaluation of the mean
value of $G^{2}$ and it was calculated before in \cite{1995}.

The second term in the expansion (\ref{des1}) contributes with the
second and the fourth terms in the expansion (\ref{des2}).

With the second term of (\ref{des2}) two topologically independent
diagrams appear which are represented in Fig \ref{G22}.

\begin{figure}[h]
\begin{center}
\includegraphics[scale=0.35,angle=0]{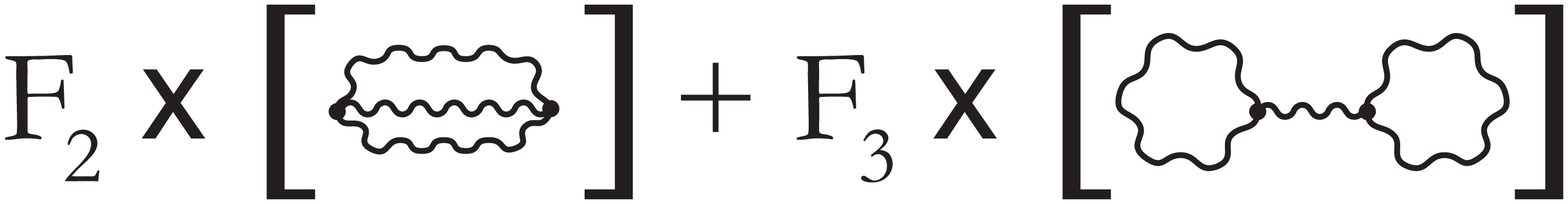}
\end{center}
\vspace{-0.7cm} \caption{Diagram 2.} \label{G22}
\end{figure}

\newpage

With the fourth term of (\ref{des2}), one diagram appears which is
represented in Fig \ref{G23}.

\begin{figure}[h]
\begin{center}
\includegraphics[scale=0.17,angle=0]{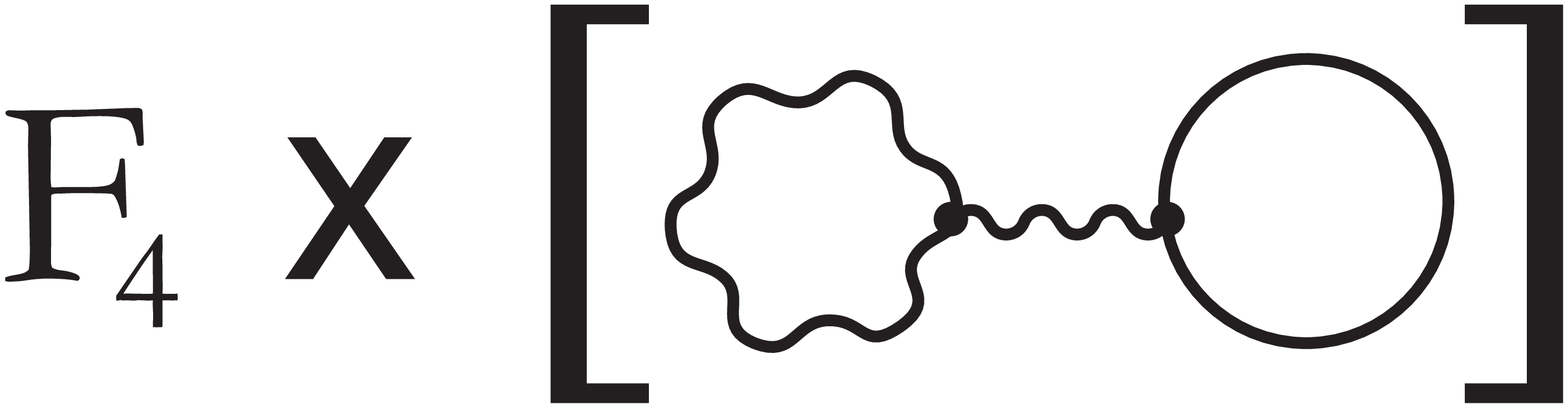}
\end{center}
\vspace{-0.7cm} \caption{Diagram 3.} \label{G23}
\end{figure}

For the calculation of the first diagram in Fig. \ref{G22}, when
it is taken into account the $i\epsilon$ prescription for the
gluon propagator, is obtained a null contribution.

The second diagram in Fig. \ref{G22} and the one in Fig. \ref{G23}
generate a null contribution to the mean value of $G^{2}$. This is
due to the appearance of terms of the type $f^{abc}\delta ^{bc}$
and $T^{ij,a}\delta ^{ij}$ respectively which are equal to zero
due to the antisymmetry of the structure constants $f^{abc}$\
(similar arguments are used for the analogous diagrams having
ghosts) and also due to the vanishing traces of the matrices
$T^{ij,a}$.

Next, the first term in the expansion (\ref{des1}), up to the
order $g^{2}$, contributes with the first, third, fifth, sixth and
seventh of the expansion (\ref{des2}), as well as with the four
terms of (\ref{des3}).

With the first term of (\ref{des2}), the diagram represented in
Fig. \ref{G24} is obtained.

\begin{figure}[h]
\begin{center}
\includegraphics[scale=0.12,angle=0]{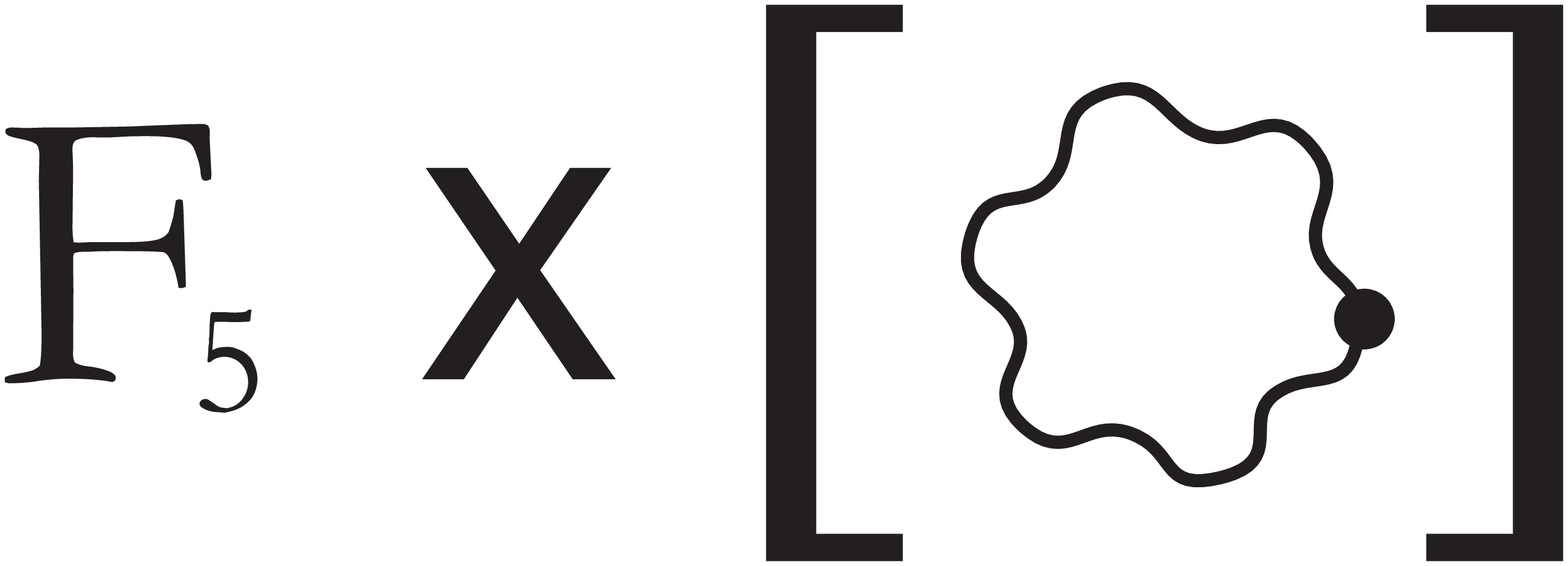}
\end{center}
\vspace{-0.7cm} \caption{Diagram 4.} \label{G24}
\end{figure}

With the third term of (\ref{des2}) and the first of (\ref{des3}),
the diagram shown in Fig. \ref{G25} is obtained. It can be
observed that the net effect of the normalization factor of the
generating functional has been to cancel out the vacuum diagrams,
the same occurs for the diagrams which follow.

\begin{figure}[h]
\begin{center}
\includegraphics[scale=0.17,angle=0]{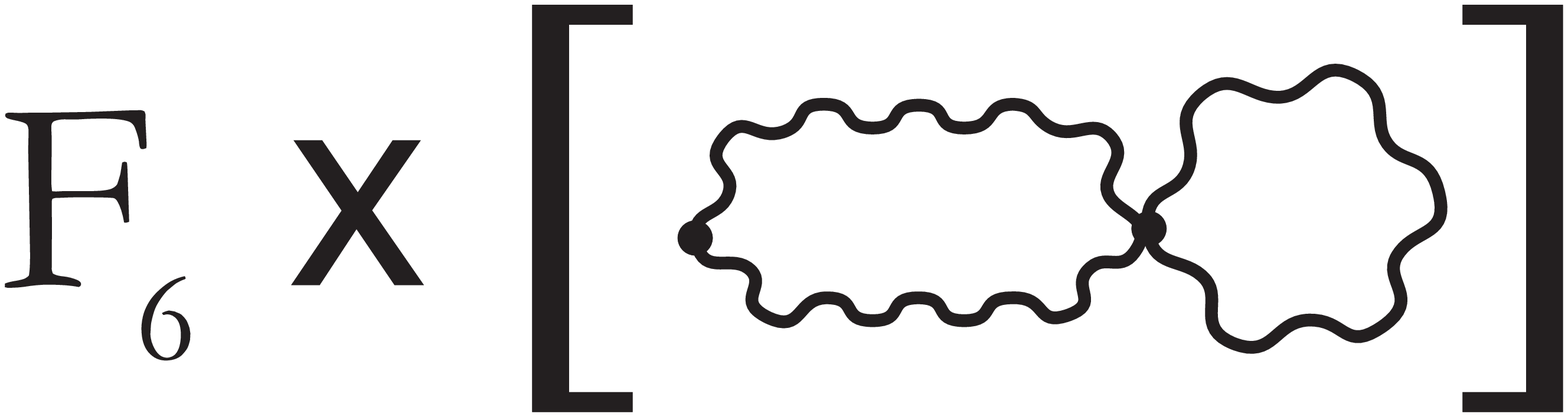}
\end{center}
\vspace{-0.7cm} \caption{Diagram 5.} \label{G25}
\end{figure}

With the fifth term of (\ref{des2}) and the second of
(\ref{des3}), the three diagrams represented in Fig. \ref{G26}
appear; with the sixth term of (\ref{des2}) and the third of
(\ref{des3}), the diagrams represented in Fig. \ref{G27} are
obtained; finally for the action of (\ref{des1}) on the seventh
term of (\ref{des2}) and the fourth of (\ref{des3}) it follows
Fig. \ref{G28}.

\newpage

\begin{figure}[h]
\begin{center}
\includegraphics[scale=0.42,angle=0]{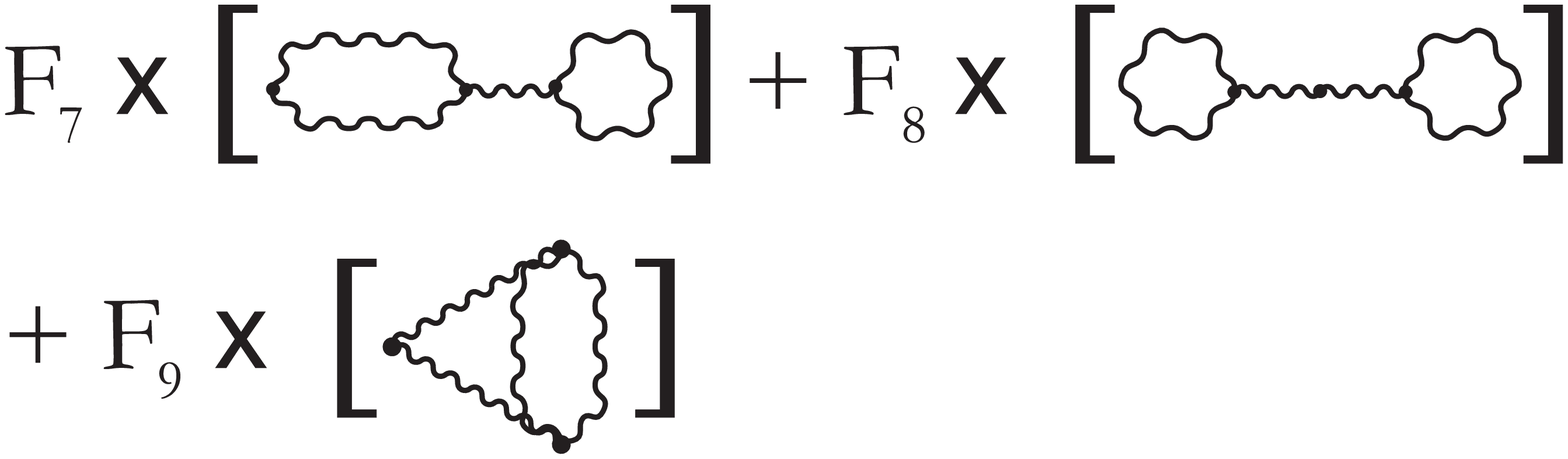}
\end{center}
\vspace{-0.7cm} \caption{Diagram 6.} \label{G26}
\end{figure}
\begin{figure}[h]
\begin{center}
\includegraphics[scale=0.38,angle=0]{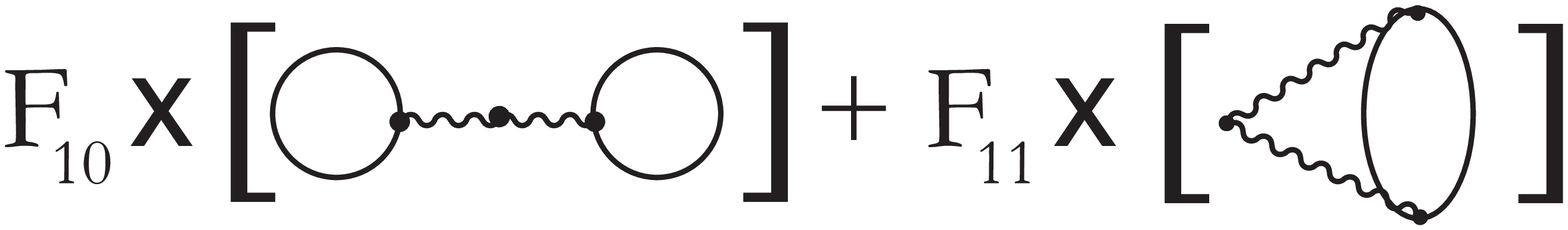}
\end{center}
\vspace{-0.7cm} \caption{Diagram for 7.} \label{G27}
\end{figure}
\begin{figure}[h]
\begin{center}
\includegraphics[scale=0.42,angle=0]{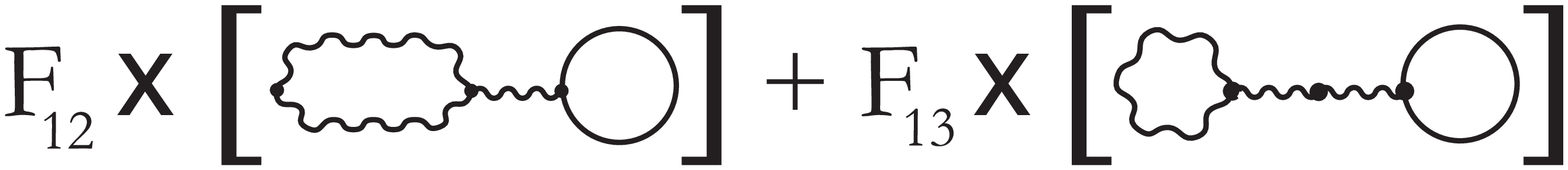}
\end{center}
\vspace{-0.7cm} \caption{Diagram 8.} \label{G28}
\end{figure}

Once, all the diagrams generated by the first term in the
expansion (\ref{des1}) were examined, the results for their
contributions are analyzed.

For the diagram represented in Fig. \ref{G24}, in the momentum
representation it is obtained:
\begin{equation}
\int \frac{d^{4}k}{\left(2\pi \right) ^{8}}\delta
^{a_{i}a_{j}}\left[ -\left(g_{\mu _{i}\mu _{j}}k^{2}-k_{\mu
_{i}}k_{\mu _{j}}\right) \right] \delta ^{a_{i}a_{j}}g^{\mu
_{i}\mu _{j}}\left(-iC\delta \left(k\right) \right) \int d^{4}x=0,
\end{equation}
Thanks to the antisymmetry property of the structure constants and
the vanishing traces of the matrices $T^{ij,a}$, the first and
second diagrams of Fig. \ref{G26}, the first of Fig. \ref{G27} and
the two of the Fig. \ref{G28} will also vanish. In this way, it is
only needed to calculate the contributions of the following
diagrams: the one in Fig. \ref{G25}, the third of Fig. \ref{G26}
and the second of Fig. \ref{G27}. After considering dimensional
regularization and the $i\epsilon$ prescription it was found a
vanishing result for the contribution of that diagrams.

It is important to remark that the same occurs for the diagrams
generated by the ghost particles. Then the only diagram
contributing to the mean value of $G^2$, in the considered
approximation, is the one in Fig. \ref{G21}.

\chapter{Longitudinal and Scalar Modes contribution}

The longitudinal and scalar modes contribution is determined by
the expression {\small {\small{\setlength\arraycolsep{0.5pt}
\begin{eqnarray}
&&\langle 0 \mid \exp \left[ C_3^{*}\left(\alpha,P\right)
B_{\vec{p} _i}^aA_{\vec{p}_i}^{L,a}\right] \exp \left\{ i\int
d^4xJ^{\mu,a}\left(x\right) \left(A_{\vec{p}
_i}^{L,a+}f_{p_i,L,\mu }^{*}\left(x\right) +\right. \right.
\nonumber \\ &&\qquad \qquad \qquad \qquad\qquad \left. \left.
+B_{\vec{p}_i}^{a+}\left[ f_{p_i,S,\mu }^{*}\left(x\right)
+(1-\alpha)\left(A^{*}\left(p_0,x_0\right) f_{p_i,L,\mu
}^{*}\left(x\right) +B_\mu ^{*}\left(p_i,x\right) \right) \right]
\right) \right\} \nonumber \\ && \quad \times \exp \left\{ i\int
d^4xJ^{\mu,a}\left(x\right) \left(B_{\vec{p}_i}^a\left[
f_{p_i,S,\mu }\left(x\right) +(1-\alpha)\left(
A\left(p_0,x_0\right) f_{p_i,L,\mu }\left(x\right) +B_\mu \left(
p_i,x\right) \right) \right] +\right. \right.\nonumber
\\ && \qquad\qquad\qquad\qquad\qquad\qquad \qquad \left. \left.
+ A_{\vec{p} _i}^{L,a}f_{p_i,L,\mu }\left(x\right)\right) \right\}
\exp \left[ C_3\left(\alpha,P\right) B_{\vec{p}_i}^{a+}A_{\vec{p}
_i}^{L,a+}\right] \mid 0\rangle.\label{A1}
\end{eqnarray}}}

Using the same methodology of the previous work \cite{PRD,tesis},
is first calculated {\small {\small{\setlength\arraycolsep{0.5pt}
\begin{eqnarray}
&&\exp \left\{ i\int d^4xJ^{\mu,a}\left(x\right)
\left(B_{\vec{p}_i}^a\left[ f_{p_i,S,\mu }\left(x\right)
+(1-\alpha)\left(A\left(p_0,x_0\right) f_{p_i,L,\mu }\left(
x\right) +B_\mu \left(p_i,x\right) \right) \right] +\right.
\right.\nonumber\\ && \qquad\qquad\qquad \qquad\qquad \quad
\qquad\qquad\qquad \left. \left. + A_{\vec{p}
_i}^{L,a}f_{p_i,L,\mu }\left(x\right)\right) \right\} \exp \left[
C_3\left(\alpha,P\right) B_{\vec{p}_i}^{a+}A_{\vec{p}
_i}^{L,a+}\right] \mid 0\rangle \nonumber \\ &&=\exp \left\{
C_3\left(\alpha,P\right) \left(B_{\vec{p}_i}^{a+}-i\int
d^4xJ^{\mu,a}\left(x\right) f_{p_i,L,\mu }\left(x\right) \right)
\right. \times \label{A2} \\ && \quad \times \left. \left(
A_{\vec{p}_i}^{L,a+}-i\int d^4xJ^{\mu,a}\left(x\right) \left[
f_{p_i,S,\mu }\left(x\right) +(1-\alpha)\left(
A\left(p_0,x_0\right) f_{p_i,L,\mu }\left(x\right) +B_\mu \left(
p_i,x\right) \right) \right] \right)\right\}\mid 0\rangle,
\nonumber
\end{eqnarray}}}\noindent in the same way can be calculated the
left part of (\ref{A1}).

Substituting the right (\ref{A2}) and left parts in (\ref{A1}),
and introducing the following notation, {\small
\begin{eqnarray}
C^{*} &\equiv &C_3^{*}\left(\alpha,P\right),\quad C\equiv
C_3\left(\alpha,P\right), \nonumber
\\ \hat{A}^{+} &\equiv &A_{\vec{p}_i}^{L,a+},\ \hat{A}\equiv
A_{\vec{p} _i}^{L,a},\text{ \ \ }\hat{B}^{+}\equiv
B_{\vec{p}_i}^{a+},\ \hat{B} \equiv B_{\vec{p}_i}^a, \nonumber \\
a_1 &\equiv &-i\int d^4xJ^{\mu,a}\left(x\right) \left[
f_{p_i,S,\mu }^{*}\left(x\right) +(1-\alpha)\left(
A^{*}\left(p_0,x_0\right) f_{p_i,L,\mu }^{*}\left(x\right) +B_\mu
^{*}\left(p_i,x\right) \right) \right], \nonumber
\\ a_2 &\equiv &-i\int d^4xJ^{\mu,a}\left(x\right) \left[
f_{p_i,S,\mu }\left(x\right) +(1-\alpha)\left(
A\left(p_0,x_0\right) f_{p_i,L,\mu }\left(x\right) +B_\mu \left(
p_i,x\right) \right) \right], \nonumber
\\ b_1 &\equiv &-i\int d^4xJ^{\mu,a}\left(x\right) f_{p_i,L,\mu
}^{*}\left(x\right), \nonumber \\ b_2 &\equiv &-i\int d^4xJ^{\mu
,a}\left(x\right) f_{p_i,L,\mu }\left(x\right), \label{nota}
\end{eqnarray}}\noindent the expression (\ref{A1}) takes the form
\begin{eqnarray}
&&\langle 0\mid \exp \left\{ C^{*}\left(\hat{A}+a_1\right) \left(
\hat{B}+b_1\right)\right\} \exp \left\{ C\left(
\hat{B}^{+}+b_2\right) \left(\hat{A}^{+}+a_2\right) \right\} \mid
0\rangle. \label{1.34}
\end{eqnarray}

Following exactly the same procedure described previously
\cite{PRD,tesis}, the following recurrence relation is obtained
for (\ref{1.34}). {\small
\begin{eqnarray}
&&\exp \left\{ C^{*}a_1b_1+C\left(C^{*}a_1-a_2\right) \left(
C^{*}b_1-b_2\right) \sum\limits_{m=0}^n\left[ \left| C\right|
^{2\left(2m\right) }+\left| C\right| ^{2\left(2m+1\right) }\right]
\right\}\label{B4} \label{A4} \\ &&\langle 0\mid \exp \left\{
C^{*}\hat{A}\hat{B}+C^{*n+1}C^{n+1}\left(\left(
C^{*}b_1-b_2\right) \hat{A}+\left(C^{*}a_1-a_2\right)
\hat{B}\right) \right\} \exp \left\{
C\hat{B}^{+}\hat{A}^{+}\right\} \mid 0\rangle.\nonumber
\end{eqnarray}}

To take the limit $n\rightarrow \infty$ it is considered $\left|
C\right|<1$, which implies the expression (\ref{A4}) takes the
form {\small{\setlength\arraycolsep{0.5pt}
\begin{eqnarray}
&&\exp \left\{ C^{*}a_1b_1+C\left(C^{*}a_1-a_2\right) \left(
C^{*}b_1-b_2\right) \frac 1{\left(1-\left| C\right| ^2\right)
}\right\} \nonumber \\ &&\quad \langle 0\mid \exp \left\{
C^{*}\hat{A}\hat{B}\right\} \exp \left\{ C
\hat{B}^{+}\hat{A}^{+}\right\} \mid 0\rangle. \label{sus}
\end{eqnarray}}}

Finally the notation (\ref{nota}) is substituted in (\ref{sus}).
After that all the functions of $\vec{p}_i$ are expanded in the
vicinity of $\vec{p}_i=0$, keeping in mind that the sources are
located in a space finite region, it is necessary to consider only
the first terms in all the expansion. Then, for the expression
(\ref{sus}), the result obtained is (\ref{LonSca}).

\end{document}